\documentclass[12pt]{revtex4}
\usepackage{epsfig,amsbsy,bm,amssymb}
\usepackage{hyperref}
\usepackage{graphicx}
\usepackage{float}
\usepackage{amsmath}

\DeclareMathOperator{\arcsinh}{arcsinh}

\newcommand{\isum}%
{\mathop{\hbox{$\displaystyle\sum\kern-13.2pt\int\kern1.5pt$}}}
\renewcommand{\r}{{\bm r}}

\newcommand{\p}{{\bm p}}

  \newcommand{\A}{{\bm A}}

  \newcommand{\e}{{\bm e}}
  \newcommand{\ve}{{\bm v}}

\newcommand{\bt}{\begin{tabular}}
\newcommand{\et}{\end{tabular}}

\newcommand{\eref}[1] {(\ref{#1})}
\newcommand{\Eref}[1] {Eq.~(\ref{#1})}

\newcommand{\Fref}[1] {Fig. \ref{#1}}
\newcommand{\Tref}[1] {Table \ref{#1}}

\newcommand{\br}{\begin{eqnarray*}}
\newcommand{\er}{\end{eqnarray*}}
\newcommand{\ba}{\begin{eqnarray}}
\newcommand{\ea}{\end{eqnarray}}
\newcommand{\be}{\begin{equation}}
\newcommand{\ee}{\end{equation}}
\newcommand{\hs}{\hspace*}

\newcommand{\rs}{\resizebox}
\newcommand{\bp}{\begin{minipage}}
\newcommand{\ep}{\end{minipage}}

\begin{document}
\thispagestyle{empty}

\bibliographystyle{apsrev}



\title{Tracking quantum clouds expansion in tunneling ionization}

\date{\today}

\author{I. A. Ivanov$^{1,2}$}
\email{igorivanov@ibs.re.kr}

\author{A.S.Kheifets$^{2}$}
\email{A.Kheifets@anu.edu.au}

\author{Kyung Taec Kim$^{1,3}$}
\email{kyungtaec@gist.ac.kr}

\affiliation{$^1$ Center for Relativistic Laser Science, Institute for
  Basic\ Science (IBS), Gwangju 61005, Republic of Korea}
	
\affiliation{$^2$ Research School of Physics, Australian National
  University, Canberra ACT 2601, Australia} 	

\affiliation{$^3$Department of Physics and Photon Science, GIST,
  Gwangju 61005, Korea}


\begin{abstract}
We study formation and evolution of the electron wave-packets in the
process of strong field ionization of various atomic targets. 
Our study is based on reformulating the problem in
terms of conditional amplitudes, i.e., the amplitudes describing
outcomes of measurements of different observables
provided that the electron is found in the ionized state after the end of the pulse. 
By choosing the electron coordinate as such an observable, we were able to 
define unambiguously the notion of the ionized wave-packets and to 
study their  formation and spread. We show 
that the evolution of the ionized wave packets obtained in this way
follows closely the classical trajectories at the initial stages of
evolution providing an {\it ab initio} quantum-mechanical
confirmation of the basic premises of the Classical Monte Carlo
Calculations approach. At the later stages of evolution the picture
becomes more complicated due to the wave packets' spread and due to
interference of wave packets originating from different field
maxima. Our approach also allowed us to obtain information about the
coordinate and velocity electron distributions at the tunnel exit.
\end{abstract}

\maketitle

\section*{Introduction}

The notion of an electron trajectory proved itself extremely useful
for the qualitative and, in many cases, quantitative description of
various ionization phenomena.  Even the simplest picture of 
classical electron motion in the field of an electromagnetic wave,
the well-known simple man model (SMM) and its predecessors
\cite{symp1,symp2,symp3,symp4,symp5,symp6,hhgd,Co94,symp7,kri,tipis,arbm}, provides a basis for understanding 
many ionization phenomena, such as above-threshold ionization (ATI)
and high harmonic generation (HHG).  The spectacular predictive power of
the SMM gave rise to a variety of techniques in which the 
ionization is described quantum-mechanically and the subsequent
electron motion is treated classically or semi-classically (classical
trajectory Monte Carlo or CTMC method)
\cite{2step,xv6,tipis,cusp3,arbm,tipis_naft,landsman2015,cmtc1}.  In
these approaches the role of quantum-mechanics (QM) consists in 
setting up initial conditions for the subsequent classical or
semi-classical electron motion, and to assigning statistical weights
to the trajectories originating at different times by employing the
notion of the instantaneous ionization rate (IIR).  Various analytical
expressions for the IIR obtained in the framework of the
quantum-mechanical strong field approximation (SFA) approach and its
subsequent developments
\cite{Keldysh64,Faisal73,Reiss80,ppt,tunr,adk1,iir} can be used for
this purpose, such as the Ammosov-Delone-Krainov (ADK)
\cite{adk1,adk}, or the Yudin-Ivanov \cite{yi} formulas.  

The initial values of the coordinates for the classical trajectories are  
determined by the tunnel exit position, which can be 
found using the field direction model
(FDM) \cite{landsman2015} or a more refined approach 
based on the use of the parabolic coordinate system \cite{bhu} 
for the atomic systems governed purely
by the Coulomb interaction.
The initial velocities in the directions perpendicular to the 
orientation of the electric field vector at the time of ionization are
typically assumed to be distributed according to the well-known SFA formula 
for the transverse velocity distribution \cite{tunr}. The initial velocity in
the direction parallel to the electric field vector at the moment of ionization is
typically assumed to be zero \cite{cmtc1}.

The success and utility of this semi-classical picture of 
atomic and molecular ionization is ultimately due to the essentially semiclassical 
nature of many aspects of the ionization phenomena which can be explained quite 
satisfactorily using semiclassical trajectory simulations. The so-called low-energy structures
in strong field ionization spectra \cite{class3,class2}, Coulomb focusing effect
\cite{class1}, nonadiabatic effects in strong field ionization \cite{class4}, frustrated tunneling ionization
\cite{rydn1}, have been studied using the trajectory-based methods. The semiclassical approaches based on the 
classical trajectories can include the interference effects as well, which can be done by using the so-called
Quantum Trajectory Monte Carlo (QTMC) approach\cite{qorb1,qtmc2}, or the
semiclassical two-step model for strong-field ionization \cite{s2step}, which allows
to obtain angular photo-electron distributions in good agreement with fully quantum calculations
based on the solution of the time-dependent Schr\"odinger
equation (TDSE). Such semiclassical simulations usually require much less computational effort than the 
fully quantum calculations and consequently, for complex targets, when numerical solution of the 
TDSE becomes unfeasible, use of such methods may provide the only possibility
to obtain quantitative description of the ionization process. 
 
If we are interested in a purely quantum mechanical description of the
motion of ionized electrons and still want to be able to use some
classical notions, one can apply the saddle-point method (SPM) 
to evaluate the SFA \cite{hhgd,tunr2,becker1,itm1,itm4,xv7,xv8}
or the Feynman's path-integral expressions for the ionization amplitude \cite{fein_ion}. 
One obtains in this
way a description of the ionization phenomena in terms of the so-called 
quantum trajectories (QT). QT are determined by the SPM equations and make the
action in the integrals determining ionization amplitudes stationary. 
QT is a generally complex electron trajectory originating at 
the complex saddle point $t_s$ and propagating till the 
final moment of time $t_f$, when the electron arrives at the
detector. This approach gives a very transparent and appealing view of the
ionization process. It is, moreover, very flexible and allows to design a number 
of different generalizations and developments \cite{qtreview,tunr2}. 
One may, for instance, 
start with the SFA ionization amplitude, ignoring effects of the ionic potential in
the continuum \cite{sfa_sem1,sfa_sem2,tunr2}, evaluate it applying the 
SPM \cite{sfa_sem2,fein_ion,carp,tunr2} and obtain a description of 
the ionization process in terms of the complex QT 
which are solutions to Newton's equations for a classical electron in the presence of the laser field.
Such a description provides a link with the SMM. To include the effects of the ionic potential
one can consider perturbatively the Coulomb effects on the  
QT used in the SFA (the so-called Coulomb corrected SFA or 
CCSFA method \cite{coul1}). Alternatively, one can consider the Coulomb and laser field effects 
on the trajectories
on equal footing and find the QT as solutions to the Newton's equations of motion 
in presence of the Coulomb and
laser fields, still using the SFA equation defining the saddle-point (the
Trajectory-based Coulomb SFA or (TCSFA) method \cite{csfa3}). Yet more generally,
one may apply the SPM to evaluate the Feynman's path-integral representation of the ionization amplitude
\cite{fein_ion}, obtaining QT which are solutions to the Newton's equations  of motion 
in presence of the Coulomb and laser fields with a more complicated 
condition defining the starting time $t_s$ of the 
trajectory.

The path connecting $t_s$ and $t_f$ in the complex time-plane is often chosen to
consist of two straight line segments: $(t_s,Re\ t_s)$ and $(Re\ t_s,t_f)$,
with $Re\ t_s$ interpreted as the tunnel exit point. 
One should bear in mind, however, that QT are all but a convenient (albeit very
useful and powerful) mathematical construct, arising as a result of
the application of the SPM for evaluation of ionization amplitudes.
In particular, the path connecting $t_s$ and $t_f$ described above is not unique 
and can be deformed to cross the
real time-axis  almost at any given point \cite{tunr2}.  This remark applies equally, of course,
to the notion of the tunnel exit used in the CTMC method \cite{tunr2,qtreview}. 

This path-dependence of the 
time and location of the tunnel exit does not affect the quantum amplitude since 
the integrals defining the amplitudes depend only on the end points $t_s$ and $t_f$ of the 
path (as long as deforming the path we do not cross singular points of the integrand in the 
complex time-plane). As it was mentioned in the review 
work \cite{tunr2}, no physical experiment can favor particular values for time or location of the
tunnel exit event, which does not prevent these notions
to be extremely useful for practical purposes.
In practice, the path connecting $t_s$ and $t_f$, consisting 
of the two straight line segments: $(t_s,Re\ t_s)$ and $(Re\ t_s,t_f)$ is the most convenient choice.
The exit time $Re\ t_s$ and the tunnel exit point are related to the corresponding sub-barrier part 
of the QT as the real part of the expression 
$\displaystyle \r=\int\limits_{t_s}^{Re\ t_s} \ve(t)\ dt$, where $\ve(t)$ is the 
complex-valued sub-barrier electron velocity. The sub-barrier part of the QT thus defined 
can be used to provide the 
necessary prerequisites for the CTMC simulations, such as position of the tunnel exit,
the transverse and longitudinal momentum distributions at the tunnel exit and the instantaneous 
ionization rate \cite{qorb1,ccpati1}. Taking the real part of the QT at $t=Re\ t_s$ as the 
tunnel exit position has the advantage that the imaginary part of the action which determines the
ionization probability is accumulated in the sub-barrier region, while subsequent propagation of the 
QT in the real time only produces phase-shift due to the change of the real part of the action. 
This choice also allows to avoid complicated issues of branching
points and branch cuts in the complex time-plane \cite{qtreview}.

Of course, all the information we can hope to obtain about a physical
system is encoded in its wave function.  Any question about the motion of
the ionized wave-packets for times within the laser pulse,
should therefore be
resolved from the solution of the TDSE, provided that this question has a physical meaning at all.
The approach based solely on the information obtained by solving the TDSE, however, 
encounters the problem of
identifying the contributions from different channels (ionization,
excitation, and so on) for times within the laser pulse, when
the wave-packets corresponding to these channels are not spatially
separated yet. The splitting of the total wave-function of the system
into the bound and ionized components seems to have been achieved in
the SFA and the Perelomov-Popov-Terent'ev (PPT) approaches
\cite{Keldysh64,Faisal73,Reiss80,ppt,tunr,adk}. Such a splitting is
not unique, however, and it is different in the two theories.
Moreover, it is not gauge-invariant in the SFA or PPT approaches
\cite{tunr2}.  Similarly, the procedure based on 
projecting out contributions of the
bound states of the field-free atomic Hamiltonian from the TDSE
wave-function, which is sometimes used to define wave-packet describing 
ionized electron, is not gauge-invariant when applied for the times within the
laser pulse duration.

In \cite{tulsky} a method, allowing to identify 
the part of the wave-function describing ionized electron and relating the TDSE-based
approach with the insight offered by the trajectory based approaches, has been proposed.
In the framework of this method the photoionized part of the wave function is singled out 
by means of applying the time dependent surface flux (tSURFF) method \cite{tserf}, which 
relies on the knowledge of the wave-function in the asymptotic region, when
the photo-ionized part of the wave function is localized in space. By applying a 
short-time filter to the ionization amplitude, calculated using the 
tSURFF method, authors were able to identify the dominant pathways which form the
photoelectron spectra.

Another group of methods allowing to connect the TDSE and the notion of trajectory is 
based on the Bohmian interpretation of the QM \cite{bohm}. 
Bohmian view of the QM introduces a well-defined notion of the electron trajectory, exactly 
reproducing at the same time all the predictions of the conventional QM \cite{bell}. 
This possibility of reintroducing trajectories in the QM framework 
has been exploited to describe ionization
of atoms \cite{bohm1DH,bohm3Dsubsycle} and molecules \cite{bohm1DH2} driven by strong laser fields, and for 
the description of the HHG process \cite{bohm1Dhhg,bhhg2,bohm3Dhhg}. An approach to the 
problem of the tunneling time, based on the Bohmian QM, has been described in 
\cite{landsman_bohm}. In \cite{bomii} the notion of the 
coordinate distributions describing ionized 
electrons has been defined using the Bohmian trajectories, which 
allowed to look at the tunnel exit problem from the perspective offered by the Bohmian QM. 

In our earlier works \cite{cori1,cori2,cori3} we described an
alternative method that allowed us to extract from the TDSE information about the
time-development of the ionization process for  times within the laser
pulse duration. The method is based on the analysis of  
two-time correlation functions, computed using the time-dependent 
wave-function describing evolution of the system, which was
obtained by solving the TDSE numerically. In essence, this procedure allows us to
use the notion of the conditional probability, where the condition is
imposed at an instant when the laser pulse is gone. In
other words, we formulate the questions about different observables
characterizing the electron motion in the following way: What would be
the probability of observing a given value of a certain observable 
during the laser pulse, provided that the electron is
found in a given state at the end of the laser pulse? We applied this
technique to study the evolution of the electron velocity distribution in
strong field ionization \cite{cori2} and to study trajectories of
the electron wave-packets for the process of the frustrated tunneling
ionization (FTI) \cite{cori3}. Here we report a study of
evolution of the ionized wave-packets for the process of strong
field ionization, based on the analysis of the information obtained from the 
numerical solution of the TDSE.

Atomic units with
$\hslash=1$, $e=1$, $m=1$ where $e$ and $m$ being the charge and the mass of the
electron are used throughout the paper.

\section*{Theory}
\label{Theory}

We consider an atom interacting with a linearly polarized laser pulse
which we define in terms of its vector potential: $\displaystyle
\bm{E}(t)=-{\partial \A(t)\over \partial t}$, where:
\be
\A(t)= -{\hat \e_z}{E_0\over\omega}\sin^2{\left({\pi t\over T_1}\right)}
\sin{(\omega t+ \phi)} \ .
\label{vp}
\ee
Here $T_1=N_cT$ is the total pulse duration and $T=2\pi/\omega$ is the
optical cycle (o.c.) corresponding to the central frequency
$\omega=0.057$~ a.u. (the wavelength of 800~nm).  In the calculations
below we use pulses with $N_c=4$. The target system is described using
the single-active electron (SAE) approximation and a spherical 
potential $V(r)$. As targets, we will consider the hydrogen atom
with $V(r)=-1/r$, a model atom with a short range (SR) Yukawa-type
potential $V(r)=-1.903 e^{-r}/r$ and the Ar atom described by means of
an effective potential \cite{oep1}. The target atom is initially in
the ground state $|\phi_0\rangle $, which is an $s$ state for the
hydrogen and Yukawa atoms (both with the ionization potential of
0.5~a.u.)  and a $p$ state with the energy -0.59 a.u. in the case of
the Ar atom.

We have shown in earlier works \cite{cori1,cori2,cori3} that tunneling
ionization dynamics can be studied in detail by analyzing suitably
chosen two-time correlation functions:
\be
C(A(t_1)B(t_2)) = 
\langle \phi_0|\hat A^H(t_1)\hat B^H(t_2)|\phi_0\rangle  \ ,
\label{c3}
\ee 
where the operators $\hat A^H(t)$ and $\hat B^H(t)$ in \eref{c3} are
taken in the Heisenberg representation, and $|\phi_0\rangle $ is the
initial state of the system. The particular choice of the operators
$\hat A$ and $\hat B$ in \Eref{c3} is dictated by the nature of the
problem under consideration. We have shown in \cite{cori2,cori3} that
by choosing for $\hat B^H(t)$ the Heisenberg form $\hat Q^H(T_1)$ of a
suitable Schr\"odinger projection operator $\hat Q$ one can study
the dynamical development of various ionization processes.
The reason why we may expect the correlation function \eref{c3} to
provide a useful dynamical information with such a choice of $\hat B$
can be easily understood if in \Eref{c3} we transform the operators to the
more familiar Schr\"odinger picture:
\begin{eqnarray}
\hat A^H(t)&=& \hat U(0,t) \hat A\hat U(t,0) \nonumber \\
\hat B^H(t)&=& \hat U(0,t) \hat B\hat U(t,0) \ ,
\label{hp}
\end{eqnarray}
where $\hat U(t,0)$ is the operator describing quantum
evolution of the system, so that the wave function of the system at
time $t$ is $\Psi(t)= \hat U(t,0)\phi_0$. Applying the transformation
\eref{hp} we rewrite \Eref{c3} as:
\be
C(A(t_1)B(t_2)) = 
\langle \hat A\Psi(t_1)|\hat U(t_1,t_2)\hat B\Psi(t_2)\rangle  \ .
\label{c31}
\ee 

Let us assume, for instance, that $\hat B=\hat P$, where $\hat P$ is
the projection operator on the continuous spectrum of the field-free
atomic Hamiltonian, and $t_2=T_1$, where $T_1$ is the moment of time
when the laser pulse is gone.  Then, according to the well-known
projection postulate of QM \cite{lampe}, the ket-vector $\hat 
P\Psi(t_2)\rangle$ represents, apart from an unimportant normalization
factor, the wave-function of the system immediately after the
measurement that has found the electron in an ionized state.  \Eref{c31} can
therefore be interpreted as a quantum-mechanical amplitude of finding
an electron in the state $|\hat A\Psi(t_1)\rangle$ at the moment
$t=t_1$ provided that the electron has been found in an ionized state after
the end of the pulse.  With a suitable choice of the operator $\hat A$
(we will discuss this choice in more detail below) we can now have a
glimpse of the dynamical characteristics of the ionized electrons for the
moments of time $t < T_1$ within the laser pulse duration.

Similarly, if we use $\hat B= \hat I- \hat P$ in
\Eref{c31} and again choose $t_2=T_1$, the expression for the correlation
function can be interpreted as an amplitude of finding an electron in
the state $|\hat A\Psi(t_1)\rangle$ at the moment $t=t_1$ provided
that the electron remains bound after the end of the pulse, which
allows us to study dynamics of the FTI process for $t < T_1$.

We can concentrate on various aspects of the electron dynamics by
choosing the operator $\hat A$ in \Eref{c31} appropriately.  We can
choose, for instance, $\hat A$ to be a projection operator in 
momentum space.  This choice together with $\hat B=\hat P$ allows us to
study the development of the ionized electron velocity distribution \cite{cori2}. 
The choice of $\hat B=
\hat I - \hat P$ and a coordinate space projection operator for $\hat
A$ allowed us to study evolution of the FTI electrons in
coordinate space. We exploit below yet another possibility, using
the following Schr\"odinger operators in the definition of the
correlation function \eref{c31}:
\ba
\hat B & = & \hat P \nonumber \\
\hat A_{z_0} & = & |\phi_{z_0}\rangle \langle \phi_{z_0}| \ ,
\label{cc}
\ea
%

where the components of the 
ket vector $|\phi_{z_0}\rangle$ in the position representation are:

\be
\phi_{z_0}(\r)= \langle \r|\phi_{z_0}\rangle = N e^{-a(\r-\e_z z_0)^2} \ .
\label{pos}
\ee

In \Eref{pos} $N$ is the
normalization factor. The ket vector $|\phi_{z_0}\rangle$ and its components in the 
coordinate basis given by \eref{pos} depend on the parameters $z_0$ and $a$,
defining a point in space with the coordinates $(0,0,z_0)$  and the resolution with
which we look at the neighborhood of this point. In the calculations below we
used  $a=4\ln2$ which gives us the spatial resolution of
approximately one atomic unit.  
We use the position representation in all the calculations below.
In this representation action of the projection operator $\hat A_{z_0}$ on a state vector 
$|\Phi\rangle$ with the components $\Phi(\r)= \langle \r|\Phi\rangle$ can be found as:

\be
\langle \r|\hat A_{z_0}|\Phi\rangle= \phi_{z_0}(\r) \int \phi^*_{z_0}(\r)\Phi(\r)\ d\r \ .
\label{pos1}
\ee

We choose $t_2=T_1$ in \Eref{c31},
where $T_1$ is duration of the laser pulse and we  will be looking
at various moments of time $t\le T_1$.  We will be thus studying
the correlation function:
\be
C(z_0,t) = 
\langle \hat A_{z_0}\Psi(t)|\hat U(t,T_1)\hat P\Psi(T_1)\rangle  \ 
\label{c32}
\ee 
for $t\le T_1$, with the operators $\hat A$ and $\hat P$ specified in
\Eref{cc}.  It is clear from the above discussion that with such a
choice, \Eref{c32} can be interpreted as giving us (apart from an
unimportant normalization factor) a quantum-mechanical amplitude of
finding the electron near the point with the coordinates $(0,0,z_0)$
at the time $t$ provided that the electron will be found in an ionized
state after the end of the pulse. In other words, this expression
provides a means of studying trajectories of the ionized electrons
during the laser pulse.

To calculate the correlation function \eref{c32} we use a procedure similar to
the one we used previously in \cite{cori1,cori2,cori3}, and we will only briefly
describe the technical details. 
The calculation can be reduced
to multiple solutions of the 3D time-dependent Schr\"odinger equation
(TDSE):
\begin{equation}
i {\partial \Psi(\r,t) \over \partial t}=
\left(\hat H_{\rm atom} + \hat H_{\rm int}(t)\right) \Psi(\r,t) \ ,
\label{tdse}
\end{equation}
where $\displaystyle H_{\rm atom}= {\hat\p^2\over 2} + V(r)$ is the
field free atomic Hamiltonian and $\hat H_{\rm int}(t)$ is the
 atom-field  interaction Hamiltonian for which we use the length form:
\be
\hat H_{\rm int}(\r,t) = \r\cdot{\bm E}(t) \ .
\label{hint} 
\ee

We first propagate the TDSE forward in time on the interval $(0,T_1)$, 
using ground atomic state as the initial state, obtaining position representation of the 
state vector $|\Psi(T_1)\rangle $. Acting on $|\Psi(T_1)\rangle $ with
the projection operator $\hat P$ we obtain the wave-function $\Phi(\r)$ corresponding to the 
vector $|\Phi\rangle= \hat P|\Psi(T_1)\rangle $. To find the vector 
$\hat U(t,T_1)\hat P\Psi(T_1)\rangle$, that we need to compute the matrix element in 
\Eref{c32}, we propagate the TDSE \eref{tdse} backward in time using $\Phi(\r)$ as
an initial (or rather final) wave-function, obtaining the vector $|\Phi(t)\rangle$
with the components $\Phi(\r,t)$ for the times $t$ within the laser pulse. 
Simultaneously, we propagate backward in time the vector $|\Psi(T_1)\rangle $,
obtaining the vector $|\Psi(t)\rangle$ and the wave-function $\Psi(\r,t)$- solution to the TDSE for 
the times $t$ within the laser pulse.
Of course, $\Psi(\r,t)$ had already been computed during the first, forward run of the TDSE, 
but we cannot store it in memory for all the times $t$ we need as it would require too much memory space. 
We recompute it again, therefore, in the process of the back-propagation of the TDSE.

Calculating overlaps of $|\Phi(t)\rangle$ and 
$|\hat A_{z0} \Psi(t)\rangle$ for a given $z_0$ and for a given set of times $t$ (we use 
the grid of $t$ with twenty points for every optical
cycle), we obtain the correlation function $C(z_0,t)$ defined in \Eref{c32}. A single
calculation using the backward propagation that we described above, gives us 
$C(z_0,t)$ for the whole grid of $t$ and a single $z_0$. The procedure is repeated 
for different values of 
$z_0$. More specifically, we used a grid of a hundred $z_0$-values equally spaced on the interval 
$(-60$ a.u., $60$ a.u.).

The TDSE was solved numerically using the procedure tested and described in detail
earlier \cite{cuspm,circ6,ndim}. The procedure relies on representing the
coordinate wave-function as a series in spherical harmonics with the
quantization axis along the laser polarization direction. 
Spherical harmonics with orders up to $L_{\rm max}=50$ were used.  The
radial variable was treated by discretizing the TDSE on a grid with
a step-size $\delta r=0.05$~a.u. in a box of size $R_{\rm
  max}=400$~a.u. The initial ground state of the system was obtained
by using a variational calculation employing the Slater basis set
\cite{ykh1} with subsequent propagation in imaginary time \cite{itp}
on the spatial grid we described above.  The necessary convergence
checks were performed. As we discussed above, to calculate the correlation function
\eref{c31} we have to propagate the TDSE both forward and backward in
time. That was achieved by using the matrix iteration method
\cite{velocity1}.

\section*{Results}
\label{Results}

\subsection*{Correlation function analysis. Short range Yukawa potential.}

\begin{figure*}[!]
\hs{-1cm}\rs{180mm} {!}{\epsffile{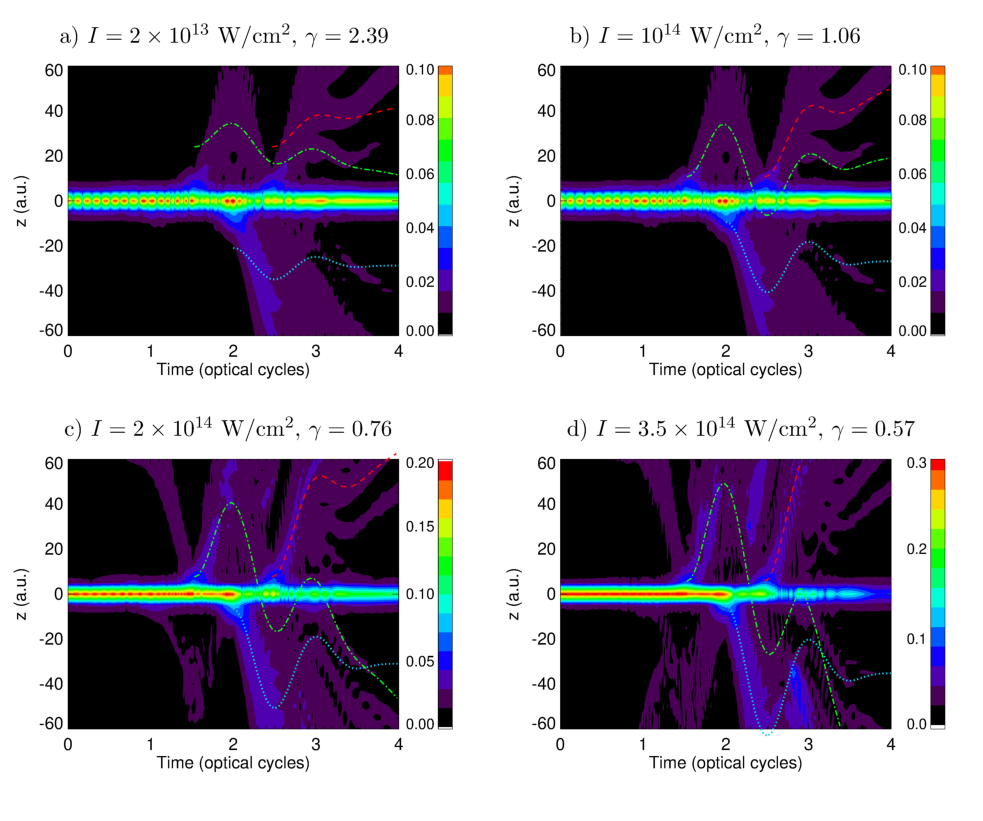}}
\caption{(Color online) Visualization of the correlation function \eref{c32} 
for the Yukawa atom at different field intensities $I$ and the CEP
  $\phi=0$. The correlation function is exponentiated ($|C(z,t)|^{1/9}$ is shown) 
for improving the visibility of the patterns. The lines in 
  the figure display the classical
  trajectories originating at the main (dots) and two auxiliary
  (dot-dash and dash) maxima of the driving laser pulse. }
\label{f1}
\end{figure*}

In this section, we present results of the correlation function
analysis based on the formal theory of the previous Section. These results
are displayed in Figures \ref{f1}-\ref{f4} below. 
Brighter colors in the figures correspond to
greater absolute values of $|C(z,t)|$.  The ionized
wave-packets spread fast (this spread is discussed in more detail below), 
so that $|C(z,t)|$ decreases very fast in magnitude when we move away from the 
instant of ionization. To see evolution of the wave-packets in greater detail 
and to be able to discern in the figures structures with very 
different magnitudes, we show 
exponentiated values $|C(z,t)|^{1/9}$.

\Fref{f1} visualizes the birth and  propagation of a photo-electron in a
model Yukawa atom with the SR potential for different field strengths
corresponding to a range of Keldysh parameters $\gamma=
\sqrt{2I_p}/A_0$ evaluated from the vector potential peak strength
$A_0$ and the ionization potential $I_0$. Our modeling covers 
both tunneling ($\gamma\lesssim 1$) and the
multiphoton ($\gamma>1$) regimes.

The lines in \Fref{f1} and other figures are used to display the
classical trajectories originating at the three main local maxima of 
the laser pulse we use. Pulse shapes for different CEPs are shown in
\Fref{f2}. Taking into account that the trajectories originating at the field maxima 
with zero velocities receive higher weights in the CTMC method, we may expect those trajectories to be 
related to the quantum picture we are analyzing. 
Following the prescriptions of the CTMC method, these classical
trajectories have been computed assuming the zero initial velocities
and the initial $z-$ coordinate defined by the FDM.  In this
model, the electron coordinate at the tunnel exit is determined as an
outer point where the electron kinetic energy becomes positive.  This
energy is obtained from energy conservation taking into account
the combined potential of the ionic core and the external
electric field which is assumed to be static.  
For brevity, we will call this construction the CTMC
trajectories.
For the Yukawa potential, the CTMC trajectories launched at the local field maxima
practically coincide with the trajectories we would have obtained had
we neglected completely any ionic potential in the classical
trajectory calculations. 
We will call below such trajectories the Coulomb-free 
trajectories. 
We do not show the Coulomb-free trajectories in the \Fref{f1}, they would be practically 
indistinguishable from the CTMC trajectories shown in the figure. 

As one can observe, except the case shown in \Fref{f1}a, the ionized
electron wave-packets initially propagate along the classical CTMC trajectories
launched at the field maxima. 
The case of the field intensity of $2\times 10^{13}$ W/cm$^2$ shown in
\Fref{f1}a stands apart. It belongs to the multiphoton regime
with the Keldysh parameter $\gamma = 2.39$.  Motion of the ionized
wave-packets, as rendered by the quantum calculation, deviates
considerably from the classical CTMC trajectories launched at the field maxima. 
As one can see from
the figure, this deviation is, in part, due to the incorrect
initial value of the initial coordinate, for which the FDM model gives too
large a value. Also, for such a value of the  
parameter $\gamma$ we cannot expect the description based on the notion of an effective potential to remain
accurate, and we cannot expect that in this ionization 
regime use of only the classical trajectories launched at the 
field maxima might provide a good approximation to the quantum picture. To obtain such an approximation 
one should, as it is done in the CMTC calculations,  
include the totality of the trajectories originating at various times within the 
laser pulse.

\begin{figure*}[h]
\hs{-1cm}\rs{180mm} {!}{\epsffile{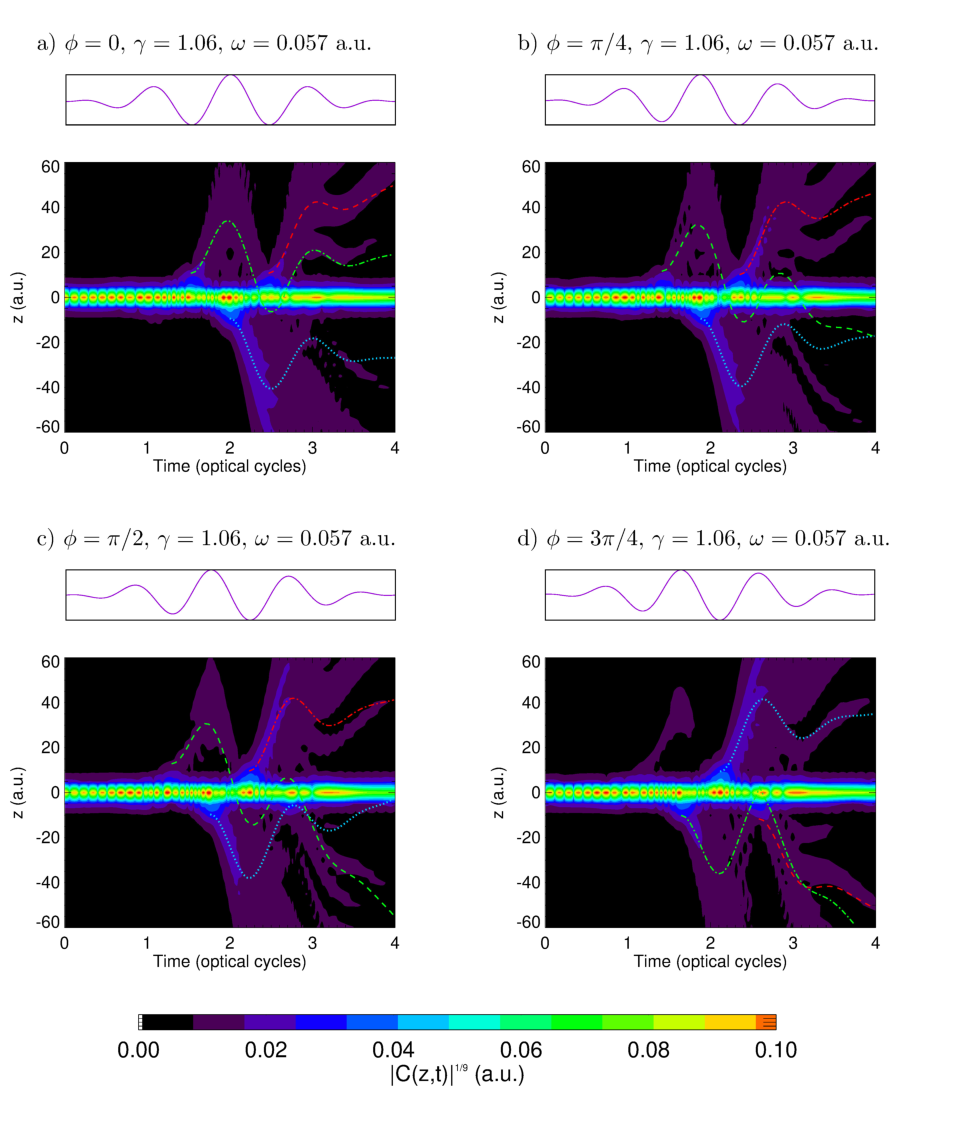}} 
\caption{(Color online) Visualization of the correlation function for
  the Yukawa atom at the field intensity $I= 10^{14}$ W/cm$^2$ and
  various CEPs $\phi$. The correlation function is exponentiated ($|C(z,t)|^{1/9}$ is shown) for improving the visibility of the patterns. 
  Lines in the figure visualize the classical trajectories
  originating at the main (dots) and two auxiliary (dot-dash and
  dash) maxima of a laser pulse.  }
\label{f2}
\end{figure*}

To relate the birth place of the photo-electron with the local maxima
of the electric field, we varied the carrier-envelope phase (CEP) of
the driving laser pulse. Results of these simulations for the Yukawa atom 
are presented in \Fref{f2}. 
We see again that initially the ionized wave-packets follow closely the
CTMC trajectories, which in turn, are practically identical to the Coulomb-free
trajectories.  These results are in accordance with the SFA in which
the birth and motion of the ionized wave packets are solely due
to the electron-laser interaction. Such an approach is fully justified
for the SR Yukawa atom. Our approach allows us to visualize how this picture
actually emerges from an {\it ab initio} TDSE calculation.

As one can see from \Fref{f1} and \Fref{f2}, at the latter stages of
evolution the ionized wave-packets broaden and their paths may deviate
considerably from the CTMC trajectories launched at the field maxima. 
This, we believe is a
consequence of the wave-packet spread and interference of the
wave-packets emitted at different times. To describe
qualitatively these effects we can use a model
based on the SFA which we describe in the Appendix.

\subsection*{Correlation function analysis. H and Ar atoms.}

\begin{figure*}[h]
\hs{-1cm}\rs{180mm}{!}{\epsffile{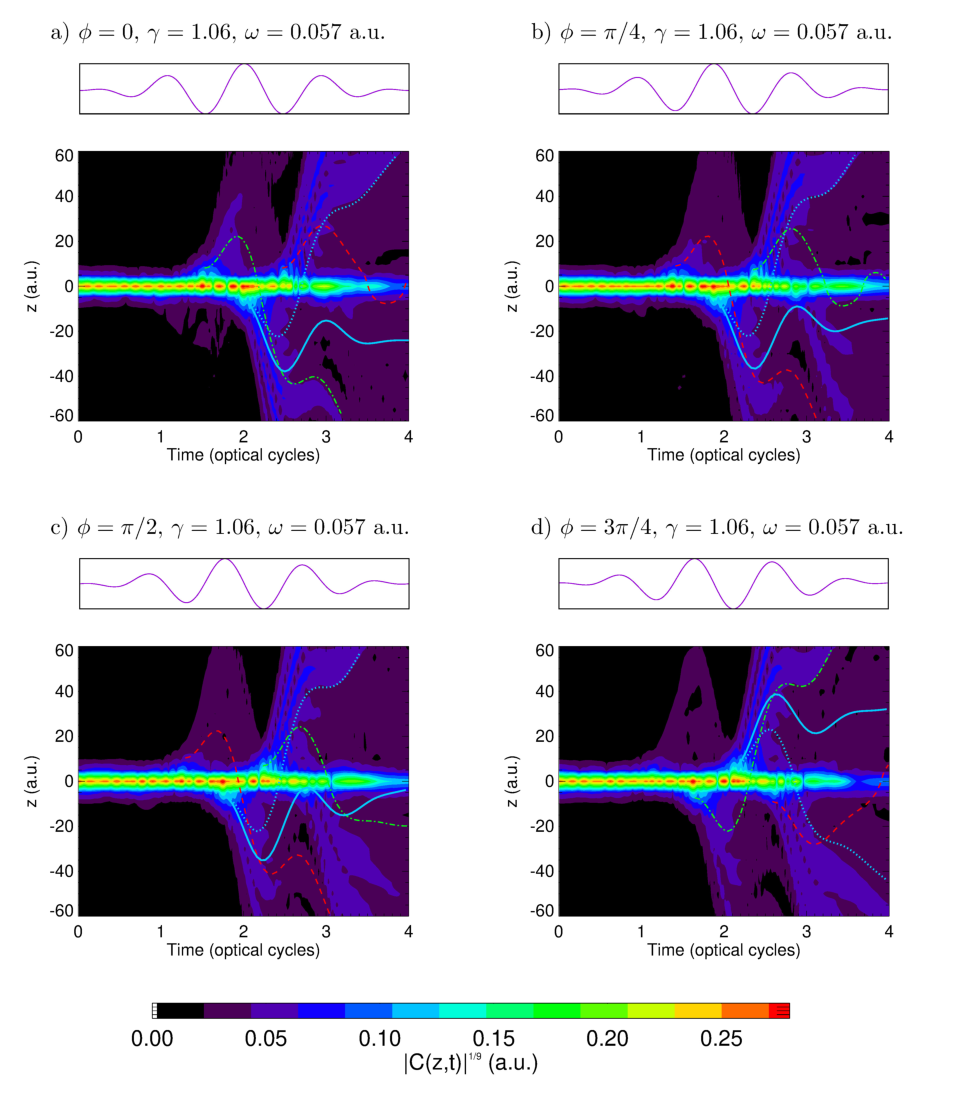}} 
\caption{(Color online) Visualization of the correlation function for
  the hydrogen atom at the field intensity $I= 10^{14}$ W/cm$^2$ and
  various CEPs $\phi$. 
  The correlation function is exponentiated ($|C(z,t)|^{1/9}$ is shown) for improving the visibility of the patterns. Lines in the figure show the classical trajectories
  originating at the main (dots) and two auxiliary (dot-dash and
  dash) maxima of a laser pulse. Solid line shows the Coulomb-free
  trajectory originating at the 
  main field maxima.}
\label{f3}
\end{figure*}

\begin{figure*}[h]
\hs{-1cm}\rs{180mm}{!}{\epsffile{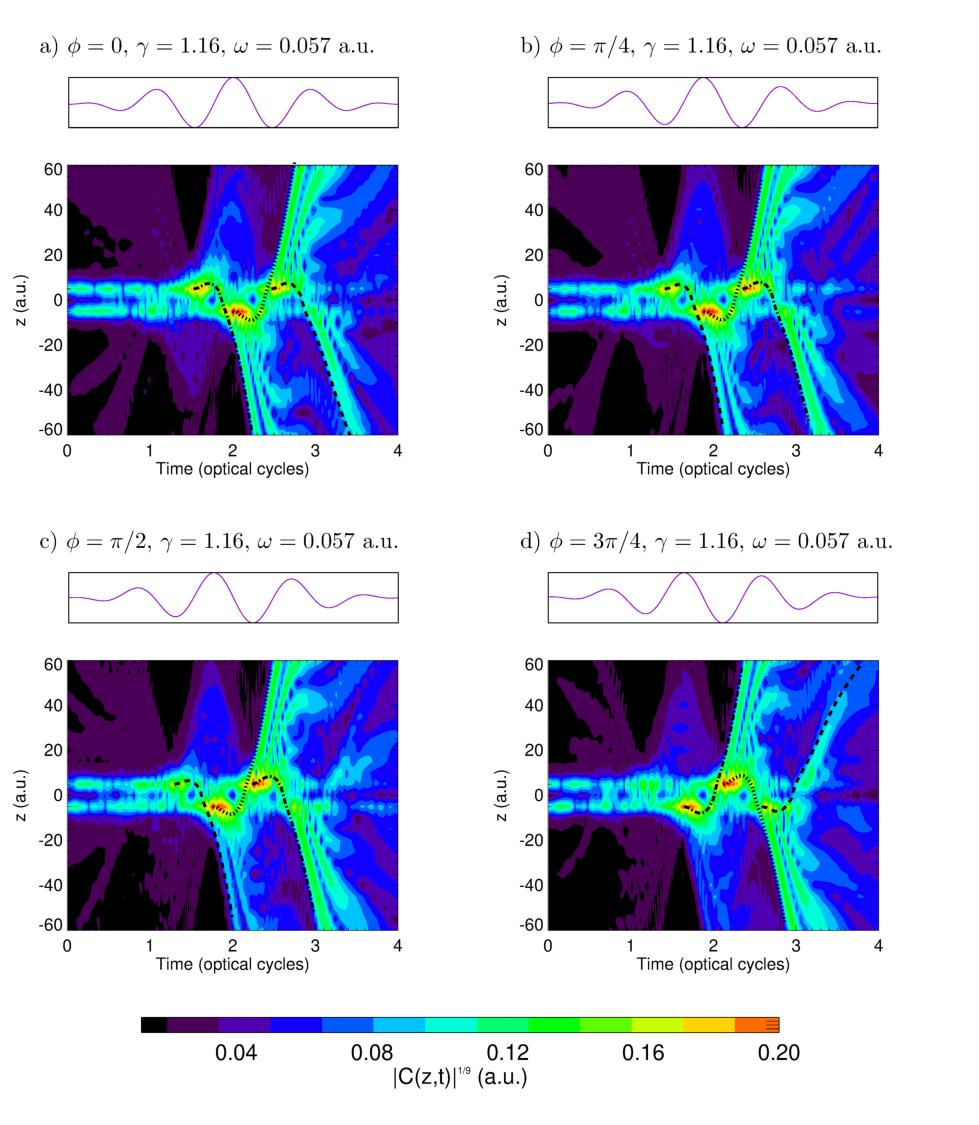}} 
\caption{(Color online) Visualization of the correlation function for
  the argon atom at the field intensity $I= 10^{14}$ W/cm$^2$, 
	initial $3p_z$ state and
  various CEPs $\phi$. The correlation function is exponentiated ($|C(z,t)|^{1/9}$ is shown) for improving 
	the visibility of the patterns. Lines in
  the figure visualize the classical trajectories originating at the main
  (dots) and two auxiliary (dot-dash and dash) maxima of a laser
  pulse.  }
\label{f4}
\end{figure*}

\Fref{f3} and \Fref{f4} are analogous to \Fref{f1} and show the
photo-electron trajectories for the hydrogen and Ar atoms,
respectively.  As in the case of the SR potential, for some time after
their birth the ionized wave-packets follow relatively closely the
classical trajectories, progressively widening. This widening and the
interference of the wave-packets born at different local maxima of the
field alters this motion at the later stages of the evolution.
Unlike the SR Yukawa
case, the long range Coulomb force introduces a considerable change
into the wave-packets dynamics. This point is illustrated in
\Fref{f3}, where the solid line shows the Coulomb-free trajectory originating
at the main field maximum, i.e., the trajectory obtained if the
ionic potential (the pure Coulomb in the hydrogen case) is
neglected. One can see that the Coulomb-free trajectory deviates quite considerably
both from the CTMC trajectory and from the TDSE correlation pattern.

\begin{figure*}[h]
\hs{-1cm}\rs{180mm}{!}{\epsffile{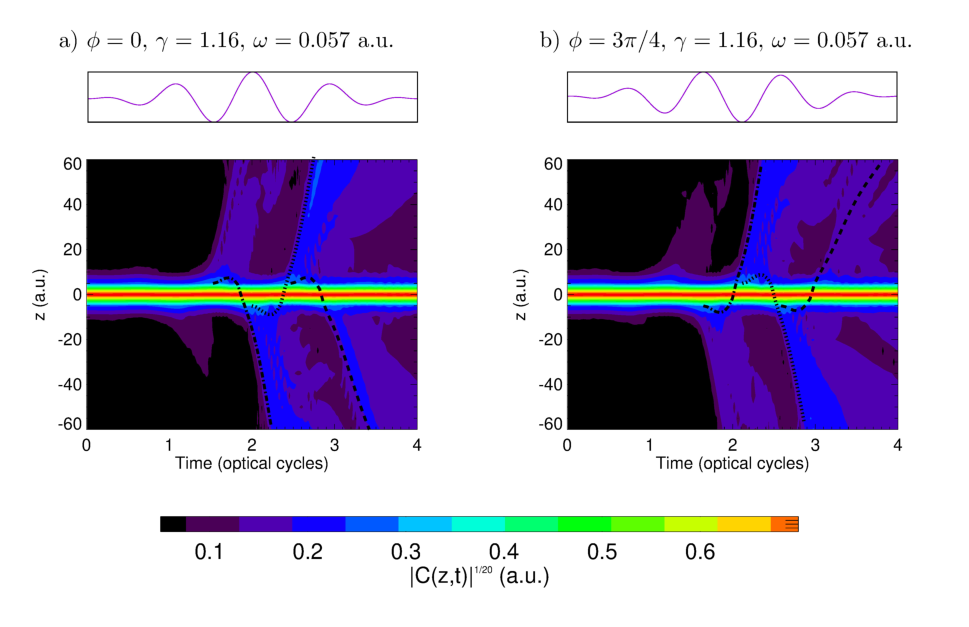}} 
\caption{(Color online) Visualization of the correlation function for
  the argon atom at the field intensity $I= 10^{14}$ W/cm$^2$, initial 
	$3p_x$ state and
  various CEPs $\phi$. 
   The correlation function is exponentiated ($|C(z,t)|^{1/20}$ is shown) for improving the visibility of the patterns. Lines in
  the figure visualize the classical trajectories originating at the main
  (dots) and two auxiliary (dot-dash and dash) maxima of a laser
  pulse.  }
\label{f41}
\end{figure*}

In \Fref{f4}  and \Fref{f41}  
we present results for the Ar atom with the initial state $3p$ orbital
oriented differently with respect to the laser polarization vector. 
\Fref{f4} shows results for the initial $3p_z$ state, oriented in $z-$ direction 
along the laser field, while \Fref{f41} shows results for the initial 
$3p_x$ state, oriented in $x$-direction perpendicular to the laser field. 
Unlike the two
previous cases of the Yukawa and hydrogen atoms, 
\Fref{f4} shows horizontal bands, present at the initial stage of the evolution
before the first maximum of the laser pulse. These bands reflect the 
distribution of the electron density along the polarization direction due to the 
nodal structure of the initial $3p_z$ state. For the cases of the Yukawa and
H atoms with the initial $s-$ state shown in \Fref{f1}, \Fref{f2} and \Fref{f3} 
we have, of course, only one  band concentrated
near the origin where the coordinate density of the unperturbed initial state is 
maximal. For the moments of time within the laser pulse, preceding the first local maximum of the 
pulse, the presence of such a band (or bands in the case of the $3p_z$ state of Ar) 
in the plots showing the correlation function
is easy to explain. It is just a consequence of the simple fact that all the ionized electrons
resided initially in the ground atomic state. According to this logic
the bands due to the correlations between the ionized electrons and the electrons 
in the initial atomic state, should disappear or diminish in brightness  
for the moments of time exceeding position of the major maximum of the field strength, when 
relatively few electrons can be ionized. In other words, if $t_m$ is the position of the 
major maximum of the pulse field strength,
then we might expect these bands to start vanishing or diminishing in
brightness for $t \gtrsim t_m$. 
We see that this is indeed the case for the correlation pattern for the ionization 
from the $3p_z$ state of argon atom shown in \Fref{f4},
where the bands describing correlations between the ionized and bound electrons vanish
for $t>t_m$. This is also the case of the correlation pattern for the hydrogen atom 
shown in \Fref{f3}, where the band around the line $z=0$ diminishes in brightness for $t> t_m$. 
The picture is apparently different for the correlation patterns for the Yukawa atom 
(\Fref{f1} and \Fref{f2}) and ionization from the $3p_x$ state of the Ar atom (\Fref{f41}).
With the exceptions of \Fref{f1}c and \Fref{f1}d showing results for the Yukawa atom for 
higher field strengths, these figures do not show any appreciable change in the degree of correlation
between ionized and bound electrons for $t>t_m$. We believe that 
this apparently counter-intuitive behavior
is an artefact which is due to a problem which is very hard to avoid in a 
numerical calculation. If we inspect the definition \eref{c32} of the correlation function, 
we see that the first step of the calculation consists in projecting out the contributions
of the bound states from the state vector  $|\Psi(T_1)\rangle$ describing the system 
at the end of the pulse. In practical calculations we perform this projection operation as follows:

\be
\hat P|\Psi(T_1)\rangle= |\Psi(T_1)\rangle- \sum\limits_{k} \langle \phi_k|\Psi(T_1)\rangle |\phi_k\rangle \ ,
\label{out}
\ee

where $|\phi_k\rangle$ describe bound states of the system. It is unavoidable in numerical
calculations that $|\phi_k\rangle$ differ slightly from the state vectors describing the true bound
atomic states. This means that after performing the projection operation, the resulting vector in 
\Eref{out} is only approximately orthogonal to all the atomic bound states vectors. Most important 
of course, is the possible non-orthogonality to the initial atomic state, which 
which would manifest itself as presence of correlations between the bound and ionized electrons 
even for the times when ionization process effectively ceases. 
The extent to which this possible non-orthogonality issue may alter the correlation pattern depends,
of course, on the magnitude of the vector $\hat P|\Psi(T_1)\rangle$. 
Clearly, this numerical problem plays more significant role when this magnitude  
$||\hat P\Psi(T_1)||^2=\langle \Psi(T_1)|\hat P |\Psi(T_1)\rangle$ is small, or, in other words, when 
the ionization probability is small. We can expect, therefore, this numerical problem to
be less important for the systems with higher ionization probabilities. This conclusion 
is confirmed by our data. Let us consider the particular case of the field intensity
of $10^{14}$ W/cm$^2$ and zero CEP. For these field parameters we obtain the following values
for the total ionization probabilities $P_{\rm ion}$ for the targets we consider: 
$P_{\rm ion}=2.49 \times 10^{-5}$ for the Yukawa atom, $P_{\rm ion}=6.61 \times 10^{-3}$ for the
hydrogen atom, $P_{\rm ion}=7.65 \times 10^{-2}$ for the Ar atom ($3p_z$ initial state), and 
$P_{\rm ion}=3.42 \times 10^{-3}$ for the Ar atom ($3p_x$ initial state). One can see that, indeed,
our data show the expected behavior of the correlation pattern, with 
the bands describing correlations between the ionized and bound electrons vanishing or
diminishing in magnitude considerably for  $t>t_m$, in the cases of higher total ionization probabilities.
viz., in the cases of the hydrogen and the Ar atom prepared initially in the $3p_z$ state.
We also see this expected behavior of the correlation patterns for the Yukawa atom 
in the cases of 
the higher field intensities shown in \Fref{f1}c and \Fref{f1}d.

\subsection*{Coordinate and velocity distributions at the moment of ionization.}

By taking the slices of the correlation patterns at $t=t_0$ along the
lines of the constant $t$ one may try to obtain some information about the distribution 
of electron coordinates at the tunnel exit.  We will be interested in a normalized
quantity:
\be d(z_0)= {|C(z,t_0)|^2\over | \langle \Psi(t_0)|\hat
  A_{z}|\Psi(t_0)\rangle|^2} \ .
\label{dz}
\ee

Here $C(z,t_0)$ is the correlation function \eref{c32}, $\hat A_z$ is
the coordinate projection operator defined in \Eref{cc} and $\Psi(t_0)$
is the solution of the TDSE describing the evolution of the system.
The normalization used in \Eref{dz} removes a trivial $z-$ dependence of the
correlation function. 

\begin{figure*}[h]
\rs{120mm} {!}{\epsffile{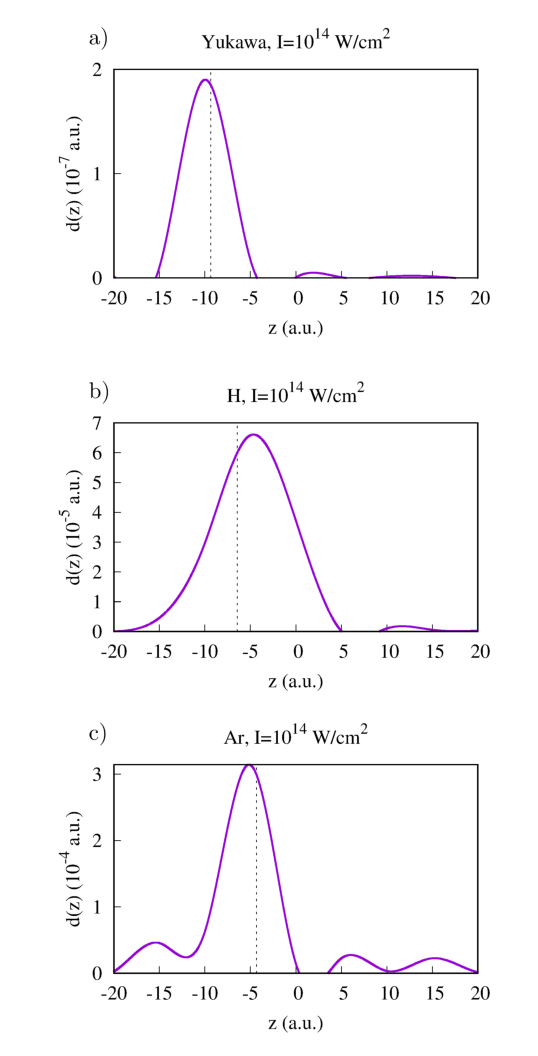}} 
\caption{(Color online) Distribution \eref{dz} for
  the ionized wave-packets originating at the main pulse maximum for
  the field intensity of $10^{14}$ W/cm$^2$ and zero CEP for the Yukawa,
	hydrogen and Ar ($3p_z$ initial state) atoms.
  Dashed vertical lines show the coordinates of the tunnel exit point given
  by the FDM model.  }
\label{f6}
\end{figure*}

In \Fref{f6} we show results for $d(z)$ by taking the slices at
$t_0=2$ o.c. for the Yukawa, hydrogen and Ar atoms at the pulse
intensity of $10^{14}$ W/cm$^2$ and zero CEP. We are thus looking
at the electrons born at the main maximum of the laser pulse.  We
should bear in mind that the correlation function is not a probability
distribution, and strictly speaking, it does not give us  
the coordinate probability distribution directly. 
We may expect, nevertheless, 
that the spatial profile of the distribution defined in \Eref{dz} may inherit 
the main features of the probability distribution, in particular, 
the position of the maximum and the width of the coordinate
probability distribution at the moment of electron ionization. These expectations are based
on the following observation. By the projection postulate 
of QM \cite{lampe}, the ket-vector 
$\displaystyle A_{z}|\Psi(t_0)\rangle/ \langle\Psi(t_0)|\hat A_{z}|\Psi(t_0)\rangle$
represents the state of the system immediately after the
measurement that detects the electron in the neighborhood of the point $(0,0,z)$ at time
$t_0$. From the \Eref{dz} and the definition \Eref{c32}, we see than that 
expression \eref{dz} can be interpreted as the probability to detect the electron in the 
ionized state $\hat P\Psi(T_1)$ at the end of the laser pulse, provided that it was found near the 
point $(0,0,z)$ at time $t_0$.

One can see that the maxima of the distributions given by \Eref{dz} 
are indeed quite close to the FDM predictions for the three targets we have considered. 
Using the plots in \Fref{f6} we can find
the full widths at half maximum (FWHM) of the coordinate
distributions. For the Yukawa and hydrogen atoms these estimates are given in \Tref{t1}. One can use
a simple check to verify if these estimates are
reasonable. 
Let us assume that the coordinate distribution is a Gaussian
with the FWHM $\Delta_z$. Then, by performing Fourier transform
we obtain the velocity distribution, which will be again a Gaussian
with the FWHM $\Delta_v$ related to the coordinate FWHM as $\Delta_v
\Delta_z= 8\ln{2}$.  We obtain in this way the estimates for the FWHMs of
the velocity distributions shown in \Tref{t1}. 
We can compare these
estimates to the FWHM following from the well-known SFA relation 
\cite{tunr} for the longitudinal electron velocity distribution:
\be W(v_z)= {\rm const} \times \exp{\left\{-2Kv_z^2(\arcsinh{\gamma}-
  \gamma(1+\gamma^2)^{-1/2})\right\} } \ .
\label{sfa}
\ee
Here $K=I_p/\omega$, $I_p$ is the target ionization potential and
$\gamma$ is Keldysh parameter. This expression gives the velocity
distribution at the detector. 
The FWHM of distribution \eref{sfa} is also
shown in \Tref{t1}. 
In general, the longitudinal velocity
distribution at the ionization instant does not need to coincide with the
distribution \eref{sfa} since this distribution may be affected by the
ionic core potential during the post-ionization propagation. 
We can, however, expect this propagation effect to play
small role for the short range Yukawa potential.  Indeed, the
FWHM of the longitudinal velocity distribution we obtain from
\Eref{sfa} agrees very well with the estimate of the velocity FWHM we
obtained above from the TDSE calculation.  The case of the Coulomb
potential is different. The coordinate distribution for the hydrogen
atom in \Fref{f6} is considerably wider than the distribution for the SR Yukawa
potential, resulting in a smaller value of around $0.5$ a.u. for the
velocity FWHM in \Tref{t1}.  Different estimates for the initial
longitudinal velocity spread for Coulomb systems can be found in the
literature, ranging from the FWHM of around $0.1$ a.u. \cite{sun1} to
$0.4$ a.u. \cite{cmtc1}. Our FWHM estimate given in \Tref{t1} seems to
agree with the latter value.

\begin{table}[h]
\caption{\label{table1} Estimates for the FWHMs of the coordinate and
  velocity distributions.}
\begin{tabular}{|c|c|c|c|}
\hline
Model & Coordinate FWHM (a.u.)  &  Velocity FWHM (a.u.), TDSE &  Velocity FWHM (a.u.), SFA\\
\hline
Yukawa  & 5.9 & 0.91 & 0.89  \\
Hydrogen & 11 & 0.50  &  \\
\hline
   \end{tabular}
\label{t1}
\end{table}

\section*{Conclusion}

In summary, we devised a procedure based on the correlation function
analysis and employed it to study ionization dynamics of the three
atomic targets: the SR Yukawa, hydrogen and Ar atoms. The starting
point of our analysis is the time-dependent wave function returned by
a numerical solution of the TDSE.  Our approach allows us to look
closely at early stages of the photo-electron evolution and to separate
various components of the wave function which, at later
stages, will contribute to distinct outcomes of the laser-atom
interaction. We achieve this result by reformulating the problem in
terms of the conditional amplitudes, i.e., the amplitudes describing
outcomes of measurements of different observables
provided that electron is found in the ionized state after the end of the pulse. 
By choosing electron coordinate as such an observable we were able to 
track the motion of the ionized electron wave-packets basing on an {\it ab initio}
TDSE calculation. 

Our study demonstrates the somewhat limited 
character of the notion of a photo-electron trajectory for the description of the ionization process.
The true photo-electron dynamics is more complex and resemble more a
``quantum cloud'' expansion. 
We demonstrate that the photo-electron wave-packets obtained in this
way follow closely the CTMC trajectories at the initial stages of the
evolution both for the SR and Coulomb systems. 
However, at the later stages of the evolution, the
picture becomes more complicated due to the spread and interference of
the wave-packets originated at different field maxima. 

In the present work 
we considered these effects using a quantum mechanical approach based on the 
correlation function analysis. Alternative description might be based
on incorporating these effects into the trajectory based methods.
Interference effects can be included in the consideration following the 
prescriptions of the QTMC method \cite{qorb1,qtmc2} or the 
semiclassical two-step model for strong-field ionization \cite{s2step},
which supply each trajectory with a phase accumulated along the trajectory and allow
thus to describe the interference effects. The dispersion effect could be described 
analogously to the description we obtained in the  SFA-based model we 
presented above, where this effect manifests itself as a spread of the wave-packet moving along the 
classical trajectory. In the simplest case given by the \Eref{wf2} the spread does not depend on the 
core potential and is described by a simple analytical formula. Such a trajectory based description
of the interference and dispersion effects possesses the advantage of the trajectory based methods,  
since it can be applied for more complex targets such as molecules,  for which use of the 
TDSE based technique becomes prohibitively computationally demanding. 

Our approach also allowed us to obtain information about coordinate and velocity 
electron distributions at the tunnel exit.

\section*{Acknowledgments}

This work was supported by the Institute for Basic Science grant (IBS-R012-D1) and 
the National Research Foundation of Korea (NRF), grant funded by the Korea government (MIST) (No. 2022R1A2C3006025). Computational works for this research were 
performed on the IBS Supercomputer Aleph in the IBS Research Solution Center.
IAI wishes to thank the Australian National University for hospitality.

\section*{Appendix: Dispersion of wave packets}

We use the well-known expression for the SFA ionization amplitude 
\cite{tunr}:
\be
a_{\p}(t)= -i\int\limits_0^t \exp{\left\{-i\int\limits_{\tau}^t {(\p+\A(u))^2\over 2} du +I_p\tau \right\} } \langle \p|\hat H_{\rm int}(\tau)\phi_0\rangle \ d\tau \ ,
\label{sfg}
\ee
where $\A(t)$ is the vector potential \eref{vp} of the pulse.

The physical meaning of the amplitude \eref{sfg1} is that it gives us the momentum 
space wave-function of the ionized wave-packet.
Fourier transform of $a_{\p}(t)$ will give us then coordinate wave-function of the ionized
wave-packet:
\be
\Psi_{ion}(\r,t)= \int e^{i\p\cdot\r} a_{\p}(t) \ d\p \ .
\label{wf}
\ee
Below, we will be interested in the absolute value $|\Psi_{ion}(\r,t)|$ of the 
coordinate wave-function \eref{wf}. Choice of the length or velocity gauges to 
describe the atom-field interaction is, therefore, immaterial for our purposes and
we use the velocity gauge in \Eref{sfg} which makes the formulas somewhat simpler.

We consider expression \eref{sfg}  for the ionization 
amplitude for times $t$ within the interval $(0,T_1)$ of the laser pulse duration. 
To evaluate  expression \eref{sfg} we employ
the SPM, supplemented with the rule used in \cite{yi} that for 
$t<T_1$ we need to consider only the saddle points $t_s$ with $Re(t_s)< t$ 
for the evaluation of the integral in \Eref{sfg}. Following the standard prescriptions
of the SPM we obtain:

\be
a_{\p}(t)=
\sum\limits_{Re(t_s)< t} (-i) e^{-{i\pi\over 4}} \sqrt{2\pi\over S''(t_s,t,\p)}
e^{-iS(t_s,t,\p)} \langle \p|\hat H_{\rm int}(t_s)\phi_0\rangle \ ,
\label{sfg1}
\ee

where $t_s$ are saddle-points of the integrand in \Eref{sfg}, satisfying 
the SPM equation $\displaystyle (\p + \A(t_s))^2+ 2I_p=0$ and:

\be
S(t_s,t,\p)= \int\limits_{t_s}^t \left({(\p+\A(u))^2\over 2} + I_p\right)\ du
\label{sfg2}
\ee

We can compute the Fourier transform defining the coordinate wave-function
\eref{wf} using
the SPM again. One can see that it is the region of small momenta $\p$ that dominate the
integral in \Eref{wf}. It is sufficient, therefore, to expand the action in \Eref{sfg2} in powers of 
$\p$ keeping only the constant, linear and quadratic terms:
\be
S(t_s,t,\p)= \alpha^s(t) p^2 + {\bm \beta}^s(t)\cdot\p+ \gamma^s(t)
\label{ex}
\ee
A simple integration than will give us for the coordinate wave-function of the 
ionized wave-packet:
\be
\Psi_{ion}(\r,t)= 
\sum\limits_{Re(t_s)< t} 
{C_s\over \alpha^s(t)^{3\over 2}}  \exp{\left\{\left(i{(\bm{\beta}^s(t)-\r)^2\over 4\alpha^s(t)}\right)-i\gamma^s(t)\right\}} \ ,
\label{wf1}
\ee
where we have absorbed all constant factors
into the factors $C_s$.

To see the physical meaning of \Eref{wf1} we have to take a closer look
at the coefficients of the expansion \eref{ex}. 
The integration path in \Eref{sfg2} can be chosen to consist of two segments:
a vertical line $(t_s,Re(t_s))$ descending on the real time axis and a
horizontal segment $(Re(t_s),t)$.  We can then represent the action in \Eref{sfg2} 
as a sum $S(t_s,t,\p) = S(t_s,Re(t_s),\p)+ S(Re(t_s),t,\p)$, where
\be
S(t_s,Re(t_s),\p) = \tilde \alpha^s_1(\p)\p^2 + \tilde{\bm \beta}^s_1(\p)\cdot\p + 
\tilde\gamma^s_1(\p) \ ,
\label{sfg3}
\ee
with
\ba
\tilde\alpha^s_1 &= & -i{Im({t_s})\over 2} \nonumber \\
\tilde{\bm \beta}^s_1 &=& \int\limits_{t_s}^{Re(t_s)} \A(u)\ du \nonumber \\
\tilde\gamma^s_1 &=&  -iIm({t_s})I_p \ ,
\label{sfg4}
\ea
and:
\be
S(Re(t_s,t),\p) = \tilde \alpha^s_2(\p)\p^2 + 
\tilde{\bm \beta}^s_2(\p)\cdot\p + \tilde\gamma^s_2(\p) \ ,
\label{sfg5}
\ee
with
\ba
\tilde\alpha^s_2 &= & {t-Re({t_s})\over 2} \nonumber \\
\tilde{\bm \beta}^s_2 &=& \int\limits_{Re(t_s)}^t \A(u)\ du \nonumber \\
\tilde\gamma^s_2 &=&  (t-Re(t_s))I_p \ .
\label{sfg66}
\ea

It is customary to refer to the action \eref{sfg3} as describing 
under-the-barrier, and \eref{sfg5} as describing post-ionization 
motion. Of course, as we mentioned in the Introduction, this division is
arbitrary to a degree, since it corresponds to a particular choice of the 
integration path in \Eref{sfg2} which is not unique.

The tilted coefficients in \Eref{sfg4} and \Eref{sfg66} 
are themselves functions of $\p$.
This dependence is due to the dependence of the saddle point position $t_s$ 
on  the momentum. To obtain the untilted quantities in \Eref{ex} we must
renormalize these coefficients, by expanding 
$t_s$ in powers of momentum components and keeping  only constant, 
linear and quadratic terms in the resulting expressions. 
Corresponding formulas become rather 
bulky and add little to understanding the physical picture, we will 
not present them here. In the actual calculation reported below 
we performed all the necessary re-expansions numerically. 

Before presenting results of this calculation we will first illustrate 
the physical meaning of \Eref{wf1} by making a few simplifying assumptions
leading to more transparent formulas.
Let us suppose first that
we can drop under-the-barrier
part of the action and substitute 
the expressions for the 
tilted coefficients from \Eref{sfg66} in \Eref{wf1}. We obtain then:
\be
\Psi_{ion}(\r,t)= 
\sum\limits_{Re(t_s)< t} 
{2^{3\over 2} C_s\over (t-Re(t_s)))^{3\over 2}}
\exp{\left\{\left(i{(\bm{\beta}^s(t)-\r)^2\over 2(
t-Re(t_s))}\right)-i\gamma^s(t)\right\}} \ ,
\label{wf2}
\ee
with ${\bm \beta}_s(t)$ given by the second of equations \eref{sfg66}.
It is not difficult to see that for each $t_s$ 
exponential factor in \Eref{wf2} describes
evolution of an electron prepared at the moment $t=Re(t_s)$ in the 
state described by a delta function $\delta(\r)$, which evolves
subsequently under the action of the laser field only. For each 
$t_s$ the corresponding evolving wave-packet is weighted by a factor 
$e^{-i\gamma_s}$. Since $\gamma_s$ is complex, the exponential function
$e^{-i\gamma_s}$ is sharply peaked around the field maximum nearest to $t_s$
(it is this factor, in fact, that gives the characteristic 
exponential dependence of the ionization probability on the field strength in
the ADK and similar formulas). 
We obtain thus a simple picture of very narrow 
electron wave-packets created at times near the local field maxima and propagating 
subsequently under the action of the laser field. 

Including the under-the-barrier part of the action makes this
picture more realistic.  Assuming that 
$\alpha_s(t)= \alpha^s_1(t)+ \alpha^s_2(t)$, where 
$\alpha^s_1(t)$ and $\alpha^s_2(t)$ are given by \Eref{sfg4} and \Eref{sfg66},
we obtain from \Eref{wf1}:

\be
\Psi_{ion}(\r,t)= 
\sum\limits_{Re(t_s)< t} 
{2^{3\over 2} C_s\over (t-Re(t_s)-iIm(t_s)))^{3\over 2}}
\exp{\left\{\left(i{(\bm{\beta}^s(t)-\r)^2\over 2(
t-Re(t_s)-iIm(t_s^{0}))}\right)-i\gamma^s(t)\right\}} \ ,
\label{wf3}
\ee
where $t_s^0$ is the zero order term in the expansion of $t_s$ in powers of 
momentum. The role that this correction plays in \Eref{wf3} is quite clear. It is easy
to see that exponential factor in \Eref{wf3} describes now the spread and motion of 
the wave-packet prepared initially in a state describe by a 
Gaussian $e^{-cr^2}$ with $c={1\over 2Im(t_s)}$,
and evolving subsequently under the action of the laser field only. Role of
the under-the-barrier part of the coefficient $\alpha$ consists, therefore, 
in giving a non-zero initial spread to the ionized wave-packet. Similarly, one 
can see, that under-the-barrier part of the coefficient $\beta$ leads to the 
non-zero initial value of the coordinate. 

\begin{figure*}[h]
\hs{-1cm}\resizebox{180mm}{!}{\epsffile{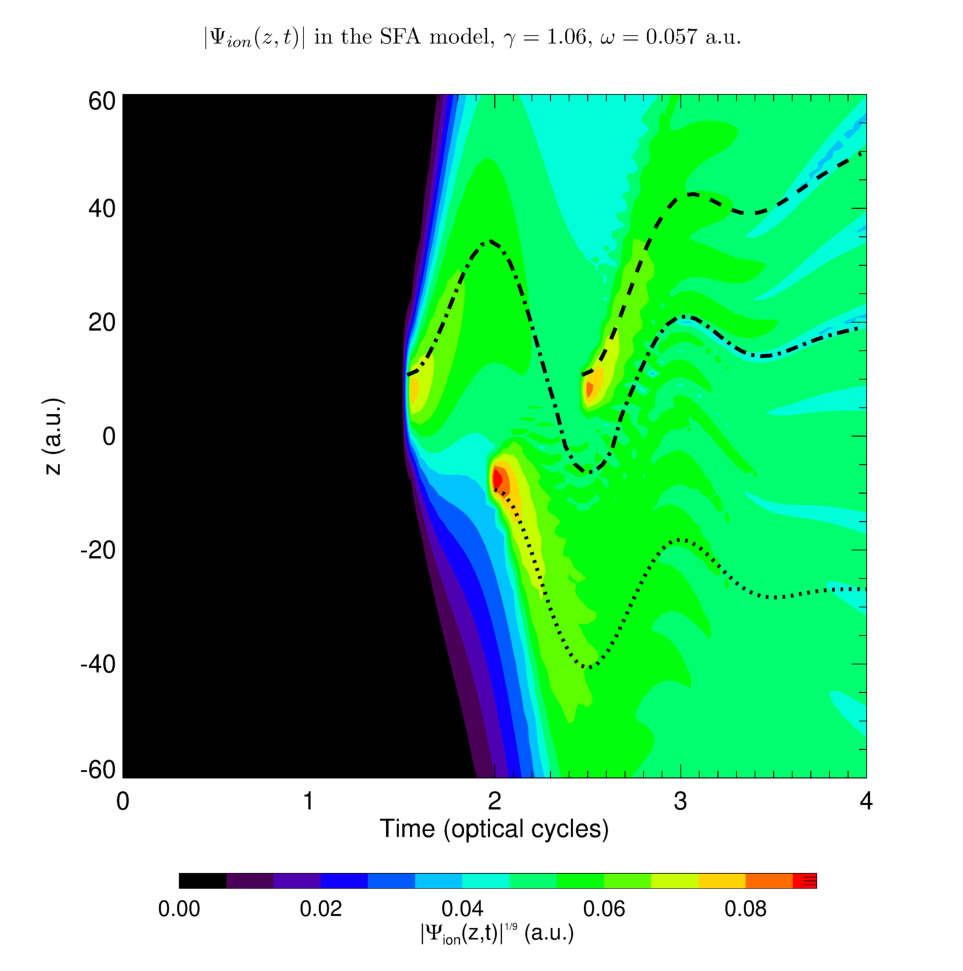}}  
\caption{(Color online) Spread of the wave-packets for the SFA coordinate
wave-function describing ionized wave-packet.
The wave-function is exponentiated ($|\Psi_{ion}(\r,t)|^{1/9}$ 
for $\r=(0,0,z)$ is shown) 
for improving the visibility of the patterns.
Lines in the figure show 
the classical trajectories originating at the main
(dots) and two auxiliary (dot-dash and dash) maxima 
of a laser pulse. 
}
\label{f5}
\end{figure*}

Results for the coordinate wave-function of the 
ionized electron provided by \Eref{wf1} by systematically
obtaining terms of the expansion \Eref{ex} from 
\Eref{sfg4} and \Eref{sfg66} are shown in
\Fref{f5}. The re-expansions needed to compute coefficients in \Eref{ex} 
from \Eref{sfg4} and \Eref{sfg66} were done numerically. 
The figure shows absolute value  $|\Psi_{ion}(\r,t)|$ along the line
$\r=(0,0,z)$ computed for the field parameters used
in \Fref{f2}a.
We note the slight offset of the $z$-coordinate of
the maxima of the correlation function at the staring points of the classical 
trajectories. This offset is due to the fact that the 
classical trajectory calculations use the FDM initial coordinates values.
The initial coordinates in the SFA calculation, on the other hand, 
are essentially determined by the second equation \eref{sfg4},
which does not include ionic potential, 
and differs, therefore, from the FDM value. One can see, nevertheless, that the
plot shown in \Fref{f5} reproduces qualitatively the main features of
the TDSE correlation pattern.  The  wave-packets follow initially the 
classical trajectories, broadening subsequently. We also see the structures 
appearing at the latest stages of evolution, for times near the end of the 
pulse, which are reminiscent of the structures seen in  \Fref{f5}.
These structures disappear if we retain in \Eref{wf1}  
only the contribution of the main field maximum, and are, therefore, 
a manifestation of the interference of 
the wave-packets born at different field maxima.

\begin{thebibliography}{74}
\expandafter\ifx\csname natexlab\endcsname\relax\def\natexlab#1{#1}\fi
\expandafter\ifx\csname bibnamefont\endcsname\relax
  \def\bibnamefont#1{#1}\fi
\expandafter\ifx\csname bibfnamefont\endcsname\relax
  \def\bibfnamefont#1{#1}\fi
\expandafter\ifx\csname citenamefont\endcsname\relax
  \def\citenamefont#1{#1}\fi
\expandafter\ifx\csname url\endcsname\relax
  \def\url#1{\texttt{#1}}\fi
\expandafter\ifx\csname urlprefix\endcsname\relax\def\urlprefix{URL }\fi
\providecommand{\bibinfo}[2]{#2}
\providecommand{\eprint}[2][]{\url{#2}}

\bibitem[{\citenamefont{Brunel}(1987)}]{symp1}
\bibinfo{author}{\bibfnamefont{F.}~\bibnamefont{Brunel}},
  \emph{\bibinfo{title}{Not-so-resonant, resonant absorption}},
  \bibinfo{journal}{Phys. Rev. Lett.} \textbf{\bibinfo{volume}{59}},
  \bibinfo{pages}{52} (\bibinfo{year}{1987}).

\bibitem[{\citenamefont{Brunel}(1990)}]{symp2}
\bibinfo{author}{\bibfnamefont{F.}~\bibnamefont{Brunel}},
  \emph{\bibinfo{title}{Harmonic generation due to plasma effects in a gas
  undergoing multiphoton ionization in the high-intensity limit}},
  \bibinfo{journal}{J. Opt. Soc. Am. B} \textbf{\bibinfo{volume}{7}},
  \bibinfo{pages}{521} (\bibinfo{year}{1990}).

\bibitem[{\citenamefont{Corkum et~al.}(1989)\citenamefont{Corkum, Burnett, and
  Brunel}}]{symp3}
\bibinfo{author}{\bibfnamefont{P.~B.} \bibnamefont{Corkum}},
  \bibinfo{author}{\bibfnamefont{N.~H.} \bibnamefont{Burnett}},
  \bibnamefont{and} \bibinfo{author}{\bibfnamefont{F.}~\bibnamefont{Brunel}},
  \emph{\bibinfo{title}{Above-threshold ionization in the long-wavelength}},
  \bibinfo{journal}{Phys.~Rev.~Lett.} \textbf{\bibinfo{volume}{62}},
  \bibinfo{pages}{1259} (\bibinfo{year}{1989}).

\bibitem[{\citenamefont{Ehlotzky}(1992)}]{symp4}
\bibinfo{author}{\bibfnamefont{F.}~\bibnamefont{Ehlotzky}},
  \emph{\bibinfo{title}{Harmonic generation in keldysh-type models}},
  \bibinfo{journal}{Nuovo Cimento D} \textbf{\bibinfo{volume}{14}},
  \bibinfo{pages}{517} (\bibinfo{year}{1992}).

\bibitem[{\citenamefont{Krause et~al.}(1992)\citenamefont{Krause, Schafer, and
  Kulander}}]{symp5}
\bibinfo{author}{\bibfnamefont{J.~L.} \bibnamefont{Krause}},
  \bibinfo{author}{\bibfnamefont{K.~J.} \bibnamefont{Schafer}},
  \bibnamefont{and} \bibinfo{author}{\bibfnamefont{K.~C.}
  \bibnamefont{Kulander}}, \emph{\bibinfo{title}{High-order harmonic generation
  from atoms and ions in the high intensity regime}},
  \bibinfo{journal}{Phys.~Rev.~Lett.} \textbf{\bibinfo{volume}{68}},
  \bibinfo{pages}{3535} (\bibinfo{year}{1992}).

\bibitem[{\citenamefont{L'Huillier et~al.}(1993)\citenamefont{L'Huillier,
  Lewenstein, Sali\`{e}res, Balcou, Ivanov, Larsson, and Wahlstr\"om}}]{symp6}
\bibinfo{author}{\bibfnamefont{A.}~\bibnamefont{L'Huillier}},
  \bibinfo{author}{\bibfnamefont{M.}~\bibnamefont{Lewenstein}},
  \bibinfo{author}{\bibfnamefont{P.}~\bibnamefont{Sali\`{e}res}},
  \bibinfo{author}{\bibfnamefont{P.}~\bibnamefont{Balcou}},
  \bibinfo{author}{\bibfnamefont{M.~Y.} \bibnamefont{Ivanov}},
  \bibinfo{author}{\bibfnamefont{J.}~\bibnamefont{Larsson}}, \bibnamefont{and}
  \bibinfo{author}{\bibfnamefont{C.~G.} \bibnamefont{Wahlstr\"om}},
  \emph{\bibinfo{title}{High-order harmonic-generation cutoff}},
  \bibinfo{journal}{Phys.~Rev.~A} \textbf{\bibinfo{volume}{48}},
  \bibinfo{pages}{R3433} (\bibinfo{year}{1993}).

\bibitem[{\citenamefont{Lewenstein et~al.}(1994)\citenamefont{Lewenstein,
  Balcou, Ivanov, L'Huillier, and Corkum}}]{hhgd}
\bibinfo{author}{\bibfnamefont{M.}~\bibnamefont{Lewenstein}},
  \bibinfo{author}{\bibfnamefont{P.}~\bibnamefont{Balcou}},
  \bibinfo{author}{\bibfnamefont{M.~Y.} \bibnamefont{Ivanov}},
  \bibinfo{author}{\bibfnamefont{A.}~\bibnamefont{L'Huillier}},
  \bibnamefont{and} \bibinfo{author}{\bibfnamefont{P.~B.}
  \bibnamefont{Corkum}}, \emph{\bibinfo{title}{Theory of high-harmonic
  generation by low-frequency laser fields}}, \bibinfo{journal}{Phys.~Rev.~A}
  \textbf{\bibinfo{volume}{49}}, \bibinfo{pages}{2117} (\bibinfo{year}{1994}).

\bibitem[{\citenamefont{Corkum}(1993)}]{Co94}
\bibinfo{author}{\bibfnamefont{P.~B.} \bibnamefont{Corkum}},
  \emph{\bibinfo{title}{Plasma perspective on strong field multiphoton
  ionization}}, \bibinfo{journal}{Phys.~Rev.~Lett.}
  \textbf{\bibinfo{volume}{71}}, \bibinfo{pages}{1994} (\bibinfo{year}{1993}).

\bibitem[{\citenamefont{Lewenstein et~al.}(1995)\citenamefont{Lewenstein,
  Kulander, Schafer, and Bucksbaum}}]{symp7}
\bibinfo{author}{\bibfnamefont{M.}~\bibnamefont{Lewenstein}},
  \bibinfo{author}{\bibfnamefont{K.~C.} \bibnamefont{Kulander}},
  \bibinfo{author}{\bibfnamefont{K.~J.} \bibnamefont{Schafer}},
  \bibnamefont{and} \bibinfo{author}{\bibfnamefont{P.~H.}
  \bibnamefont{Bucksbaum}}, \emph{\bibinfo{title}{Rings in above-threshold
  ionization: a quasiclassical analysis}}, \bibinfo{journal}{Phys.~Rev.~A}
  \textbf{\bibinfo{volume}{51}}, \bibinfo{pages}{1495} (\bibinfo{year}{1995}).

\bibitem[{\citenamefont{Krausz and Ivanov}(2009)}]{kri}
\bibinfo{author}{\bibfnamefont{F.}~\bibnamefont{Krausz}} \bibnamefont{and}
  \bibinfo{author}{\bibfnamefont{M.}~\bibnamefont{Ivanov}},
  \emph{\bibinfo{title}{Attosecond physics}}, \bibinfo{journal}{Rev. Mod.
  Phys.} \textbf{\bibinfo{volume}{81}}, \bibinfo{pages}{163}
  (\bibinfo{year}{2009}).

\bibitem[{\citenamefont{Shvetsov-Shilovski
  et~al.}(2012)\citenamefont{Shvetsov-Shilovski, Dimitrovski, and
  Madsen}}]{tipis}
\bibinfo{author}{\bibfnamefont{N.~I.} \bibnamefont{Shvetsov-Shilovski}},
  \bibinfo{author}{\bibfnamefont{D.}~\bibnamefont{Dimitrovski}},
  \bibnamefont{and} \bibinfo{author}{\bibfnamefont{L.~B.}
  \bibnamefont{Madsen}}, \emph{\bibinfo{title}{Ionization in elliptically
  polarized pulses: Multielectron polarization effects and asymmetry of
  photoelectron momentum distributions}}, \bibinfo{journal}{Phys. Rev. A}
  \textbf{\bibinfo{volume}{85}}, \bibinfo{pages}{023428}
  (\bibinfo{year}{2012}).

\bibitem[{\citenamefont{Arbo et~al.}(2015)\citenamefont{Arbo, Lemell, Nagele,
  Camus, Fechner, Krupp, Pfeifer, Lopez, Moshammer, and Burgdorfer}}]{arbm}
\bibinfo{author}{\bibfnamefont{D.~G.} \bibnamefont{Arbo}},
  \bibinfo{author}{\bibfnamefont{C.}~\bibnamefont{Lemell}},
  \bibinfo{author}{\bibfnamefont{S.}~\bibnamefont{Nagele}},
  \bibinfo{author}{\bibfnamefont{N.}~\bibnamefont{Camus}},
  \bibinfo{author}{\bibfnamefont{L.}~\bibnamefont{Fechner}},
  \bibinfo{author}{\bibfnamefont{A.}~\bibnamefont{Krupp}},
  \bibinfo{author}{\bibfnamefont{T.}~\bibnamefont{Pfeifer}},
  \bibinfo{author}{\bibfnamefont{S.~D.} \bibnamefont{Lopez}},
  \bibinfo{author}{\bibfnamefont{R.}~\bibnamefont{Moshammer}},
  \bibnamefont{and}
  \bibinfo{author}{\bibfnamefont{J.}~\bibnamefont{Burgdorfer}},
  \emph{\bibinfo{title}{Ionization of argon by two-color laser pulses with
  coherent phase control}}, \bibinfo{journal}{Phys.~Rev.~A}
  \textbf{\bibinfo{volume}{92}}, \bibinfo{pages}{023402}
  (\bibinfo{year}{2015}).

\bibitem[{\citenamefont{Shvetsov-Shilovski}(2021)}]{2step}
\bibinfo{author}{\bibfnamefont{N.~I.} \bibnamefont{Shvetsov-Shilovski}},
  \emph{\bibinfo{title}{Semiclassical two-step model for ionization by a strong
  laser pulse: further developments and applications}}, \bibinfo{journal}{Eur.
  Phys. J. D} \textbf{\bibinfo{volume}{75}}, \bibinfo{pages}{130}
  (\bibinfo{year}{2021}).

\bibitem[{\citenamefont{{Hofmann, Cornelia}
  et~al.}(2021)\citenamefont{{Hofmann, Cornelia}, {Bray, Alexander}, {Koch,
  Werner}, {Ni, Hongcheng}, and {Shvetsov-Shilovski, Nikolay I.}}}]{xv6}
\bibinfo{author}{\bibnamefont{{Hofmann, Cornelia}}},
  \bibinfo{author}{\bibnamefont{{Bray, Alexander}}},
  \bibinfo{author}{\bibnamefont{{Koch, Werner}}},
  \bibinfo{author}{\bibnamefont{{Ni, Hongcheng}}}, \bibnamefont{and}
  \bibinfo{author}{\bibnamefont{{Shvetsov-Shilovski, Nikolay I.}}},
  \emph{\bibinfo{title}{Quantum battles in attoscience: tunnelling}},
  \bibinfo{journal}{Eur. Phys. J. D}
  \textbf{\bibinfo{volume}{75}}(\bibinfo{number}{7}), \bibinfo{pages}{208}
  (\bibinfo{year}{2021}).

\bibitem[{\citenamefont{Pfeiffer et~al.}(2012)\citenamefont{Pfeiffer, Cirelli,
  Landsman, Smolarski, Dimitrovski, Madsen, and Keller}}]{cusp3}
\bibinfo{author}{\bibfnamefont{A.~N.} \bibnamefont{Pfeiffer}},
  \bibinfo{author}{\bibfnamefont{C.}~\bibnamefont{Cirelli}},
  \bibinfo{author}{\bibfnamefont{A.~S.} \bibnamefont{Landsman}},
  \bibinfo{author}{\bibfnamefont{M.}~\bibnamefont{Smolarski}},
  \bibinfo{author}{\bibfnamefont{D.}~\bibnamefont{Dimitrovski}},
  \bibinfo{author}{\bibfnamefont{L.~B.} \bibnamefont{Madsen}},
  \bibnamefont{and} \bibinfo{author}{\bibfnamefont{U.}~\bibnamefont{Keller}},
  \emph{\bibinfo{title}{Probing the longitudinal momentum spread of the
  electron wave packet at the tunnel exit}}, \bibinfo{journal}{Phys. Rev.
  Lett.} \textbf{\bibinfo{volume}{109}}, \bibinfo{pages}{083002}
  (\bibinfo{year}{2012}).

\bibitem[{\citenamefont{Dimitrovski and Madsen}(2015)}]{tipis_naft}
\bibinfo{author}{\bibfnamefont{D.}~\bibnamefont{Dimitrovski}} \bibnamefont{and}
  \bibinfo{author}{\bibfnamefont{L.~B.} \bibnamefont{Madsen}},
  \emph{\bibinfo{title}{Theory of low-energy photoelectrons in strong-field
  ionization by laser pulses with large ellipticity}}, \bibinfo{journal}{Phys.
  Rev. A} \textbf{\bibinfo{volume}{91}}, \bibinfo{pages}{033409}
  (\bibinfo{year}{2015}).

\bibitem[{\citenamefont{Landsman and Keller}(2015)}]{landsman2015}
\bibinfo{author}{\bibfnamefont{A.~S.} \bibnamefont{Landsman}} \bibnamefont{and}
  \bibinfo{author}{\bibfnamefont{U.}~\bibnamefont{Keller}},
  \emph{\bibinfo{title}{Attosecond science and the tunnelling time problem}},
  \bibinfo{journal}{Physics Reports} \textbf{\bibinfo{volume}{547}},
  \bibinfo{pages}{1} (\bibinfo{year}{2015}).

\bibitem[{\citenamefont{Hofmann et~al.}(2013)\citenamefont{Hofmann, Landsman,
  Cirelli, Pfeiffer, and Keller}}]{cmtc1}
\bibinfo{author}{\bibfnamefont{C.}~\bibnamefont{Hofmann}},
  \bibinfo{author}{\bibfnamefont{A.}~\bibnamefont{Landsman}},
  \bibinfo{author}{\bibfnamefont{C.}~\bibnamefont{Cirelli}},
  \bibinfo{author}{\bibfnamefont{A.}~\bibnamefont{Pfeiffer}}, \bibnamefont{and}
  \bibinfo{author}{\bibfnamefont{U.}~\bibnamefont{Keller}},
  \emph{\bibinfo{title}{Comparison of different approaches to the longitudinal
  momentum spread after tunnel ionization}}, \bibinfo{journal}{J.~Phys.~B}
  \textbf{\bibinfo{volume}{46}}, \bibinfo{pages}{125601}
  (\bibinfo{year}{2013}).

\bibitem[{\citenamefont{Keldysh}(1965)}]{Keldysh64}
\bibinfo{author}{\bibfnamefont{L.~V.} \bibnamefont{Keldysh}},
  \emph{\bibinfo{title}{Ionization in the field of a strong electromagnetic
  wave}}, \bibinfo{journal}{Sov. Phys. -JETP} \textbf{\bibinfo{volume}{20}},
  \bibinfo{pages}{1307} (\bibinfo{year}{1965}).

\bibitem[{\citenamefont{Faisal}(1973)}]{Faisal73}
\bibinfo{author}{\bibfnamefont{F.~H.~M.} \bibnamefont{Faisal}},
  \emph{\bibinfo{title}{Multiple absorption of laser photons by atoms}},
  \bibinfo{journal}{J.~Phys.~B} \textbf{\bibinfo{volume}{6}},
  \bibinfo{pages}{L89} (\bibinfo{year}{1973}).

\bibitem[{\citenamefont{Reiss}(1980)}]{Reiss80}
\bibinfo{author}{\bibfnamefont{H.~R.} \bibnamefont{Reiss}},
  \emph{\bibinfo{title}{Effect of an intense electromagnetic field on a weakly
  bound system}}, \bibinfo{journal}{Phys.~Rev.~A}
  \textbf{\bibinfo{volume}{22}}, \bibinfo{pages}{1786} (\bibinfo{year}{1980}).

\bibitem[{\citenamefont{Perelomov et~al.}(1966)\citenamefont{Perelomov, Popov,
  and Terentiev}}]{ppt}
\bibinfo{author}{\bibfnamefont{A.~M.} \bibnamefont{Perelomov}},
  \bibinfo{author}{\bibfnamefont{V.~S.} \bibnamefont{Popov}}, \bibnamefont{and}
  \bibinfo{author}{\bibfnamefont{M.~V.} \bibnamefont{Terentiev}},
  \emph{\bibinfo{title}{Ionization of atoms in an alternating electric field}},
  \bibinfo{journal}{Sov. Phys. -JETP} \textbf{\bibinfo{volume}{23}},
  \bibinfo{pages}{924} (\bibinfo{year}{1966}).

\bibitem[{\citenamefont{Popov}(2004)}]{tunr}
\bibinfo{author}{\bibfnamefont{V.~S.} \bibnamefont{Popov}},
  \emph{\bibinfo{title}{Tunnel and multiphoton ionization of atoms and ions in
  a strong laser field}}, \bibinfo{journal}{Physics-Uspekhi}
  \textbf{\bibinfo{volume}{47}}, \bibinfo{pages}{855} (\bibinfo{year}{2004}).

\bibitem[{\citenamefont{Ammosov et~al.}(1986)\citenamefont{Ammosov, Delone, and
  Krainov}}]{adk1}
\bibinfo{author}{\bibfnamefont{M.~V.} \bibnamefont{Ammosov}},
  \bibinfo{author}{\bibfnamefont{N.~B.} \bibnamefont{Delone}},
  \bibnamefont{and} \bibinfo{author}{\bibfnamefont{V.~P.}
  \bibnamefont{Krainov}}, \emph{\bibinfo{title}{Tunnel ionization of complex
  atoms and of atomic ions in an alternating electromagnetic field}},
  \bibinfo{journal}{Sov. Phys. -JETP} \textbf{\bibinfo{volume}{64}},
  \bibinfo{pages}{1191} (\bibinfo{year}{1986}).

\bibitem[{\citenamefont{Ivanov et~al.}(2018)\citenamefont{Ivanov, Hofmann,
  Ortmann, Landsman, Nam, and Kim}}]{iir}
\bibinfo{author}{\bibfnamefont{I.}~\bibnamefont{Ivanov}},
  \bibinfo{author}{\bibfnamefont{C.}~\bibnamefont{Hofmann}},
  \bibinfo{author}{\bibfnamefont{L.}~\bibnamefont{Ortmann}},
  \bibinfo{author}{\bibfnamefont{A.~S.} \bibnamefont{Landsman}},
  \bibinfo{author}{\bibfnamefont{C.~H.} \bibnamefont{Nam}}, \bibnamefont{and}
  \bibinfo{author}{\bibfnamefont{K.~T.} \bibnamefont{Kim}},
  \emph{\bibinfo{title}{Instantaneous ionization rate as a functional
  derivative}}, \bibinfo{journal}{Communications Physics}
  \textbf{\bibinfo{volume}{1}}, \bibinfo{pages}{81} (\bibinfo{year}{2018}).

\bibitem[{\citenamefont{Delone and Krainov}(1991)}]{adk}
\bibinfo{author}{\bibfnamefont{N.~B.} \bibnamefont{Delone}} \bibnamefont{and}
  \bibinfo{author}{\bibfnamefont{V.~P.} \bibnamefont{Krainov}},
  \emph{\bibinfo{title}{Energy and angular electron spectra for the tunnel
  ionization of atoms by strong low-frequency radiation}}, \bibinfo{journal}{J.
  Opt. Soc. Am. B} \textbf{\bibinfo{volume}{8}}, \bibinfo{pages}{1207}
  (\bibinfo{year}{1991}).

\bibitem[{\citenamefont{Yudin and Ivanov}(2001)}]{yi}
\bibinfo{author}{\bibfnamefont{G.~L.} \bibnamefont{Yudin}} \bibnamefont{and}
  \bibinfo{author}{\bibfnamefont{M.~Y.} \bibnamefont{Ivanov}},
  \emph{\bibinfo{title}{Nonadiabatic tunnel ionization: Looking inside a laser
  cycle}}, \bibinfo{journal}{Phys. Rev. A} \textbf{\bibinfo{volume}{64}},
  \bibinfo{pages}{013409} (\bibinfo{year}{2001}).

\bibitem[{\citenamefont{Hu et~al.}(1997)\citenamefont{Hu, Liu, and Chen}}]{bhu}
\bibinfo{author}{\bibfnamefont{B.}~\bibnamefont{Hu}},
  \bibinfo{author}{\bibfnamefont{J.}~\bibnamefont{Liu}}, \bibnamefont{and}
  \bibinfo{author}{\bibfnamefont{S.~G.} \bibnamefont{Chen}},
  \emph{\bibinfo{title}{Plateau in above-threshold-ionization spectra and
  chaotic behavior in rescattering processes}}, \bibinfo{journal}{Phys. Lett.
  A} \textbf{\bibinfo{volume}{236}}, \bibinfo{pages}{533}
  (\bibinfo{year}{1997}).

\bibitem[{\citenamefont{W.Becker et~al.}(2014)\citenamefont{W.Becker,
  Goreslavski, Milo\u{s}evi\'c, and Paulus}}]{class3}
\bibinfo{author}{\bibnamefont{W.Becker}}, \bibinfo{author}{\bibfnamefont{S.~P.}
  \bibnamefont{Goreslavski}}, \bibinfo{author}{\bibfnamefont{D.~B.}
  \bibnamefont{Milo\u{s}evi\'c}}, \bibnamefont{and}
  \bibinfo{author}{\bibfnamefont{G.~G.} \bibnamefont{Paulus}},
  \emph{\bibinfo{title}{Low-energy electron rescattering in laser-induced
  ionization}}, \bibinfo{journal}{J.~Phys.~B} \textbf{\bibinfo{volume}{47}},
  \bibinfo{pages}{204022} (\bibinfo{year}{2014}).

\bibitem[{\citenamefont{Liu and Hatsagortsyan}(2010)}]{class2}
\bibinfo{author}{\bibfnamefont{C.}~\bibnamefont{Liu}} \bibnamefont{and}
  \bibinfo{author}{\bibfnamefont{K.~Z.} \bibnamefont{Hatsagortsyan}},
  \emph{\bibinfo{title}{Origin of unexpected low energy structure in
  photoelectron spectra induced by midinfrared strong laser fields}},
  \bibinfo{journal}{Phys. Rev. Lett.} \textbf{\bibinfo{volume}{105}},
  \bibinfo{pages}{113003} (\bibinfo{year}{2010}).

\bibitem[{\citenamefont{Brabec et~al.}(1996)\citenamefont{Brabec, Ivanov, and
  Corkum}}]{class1}
\bibinfo{author}{\bibfnamefont{T.}~\bibnamefont{Brabec}},
  \bibinfo{author}{\bibfnamefont{M.~Y.} \bibnamefont{Ivanov}},
  \bibnamefont{and} \bibinfo{author}{\bibfnamefont{P.~B.}
  \bibnamefont{Corkum}}, \emph{\bibinfo{title}{Coulomb focusing in intense
  field atomic processes}}, \bibinfo{journal}{Phys. Rev. A}
  \textbf{\bibinfo{volume}{54}}, \bibinfo{pages}{R2551} (\bibinfo{year}{1996}).

\bibitem[{\citenamefont{Hofmann et~al.}(2014)\citenamefont{Hofmann, Landsman,
  Zielinski, Cirelli, Zimmermann, Scrinzi, and Keller}}]{class4}
\bibinfo{author}{\bibfnamefont{C.}~\bibnamefont{Hofmann}},
  \bibinfo{author}{\bibfnamefont{A.~S.} \bibnamefont{Landsman}},
  \bibinfo{author}{\bibfnamefont{A.}~\bibnamefont{Zielinski}},
  \bibinfo{author}{\bibfnamefont{C.}~\bibnamefont{Cirelli}},
  \bibinfo{author}{\bibfnamefont{T.}~\bibnamefont{Zimmermann}},
  \bibinfo{author}{\bibfnamefont{A.}~\bibnamefont{Scrinzi}}, \bibnamefont{and}
  \bibinfo{author}{\bibfnamefont{U.}~\bibnamefont{Keller}},
  \emph{\bibinfo{title}{Interpreting electron-momentum distributions and
  nonadiabaticity in strong-field ionization}}, \bibinfo{journal}{Phys. Rev. A}
  \textbf{\bibinfo{volume}{90}}, \bibinfo{pages}{043406}
  (\bibinfo{year}{2014}).

\bibitem[{\citenamefont{Ortmann et~al.}(2021)\citenamefont{Ortmann, Hofmann, Ivanov,
  and Landsman}}]{rydn1}
\bibinfo{author}{\bibfnamefont{L.}~\bibnamefont{Ortmann}},
  \bibinfo{author}{\bibfnamefont{C.}~\bibnamefont{Hofmann}},
  \bibinfo{author}{\bibfnamefont{I.~A.} \bibnamefont{Ivanov}},
  \bibnamefont{and} \bibinfo{author}{\bibfnamefont{A.~S.}
  \bibnamefont{Landsman}}, \emph{\bibinfo{title}{Controlling quantum numbers
  and light emission of rydberg states via the laser pulse duration}},
  \bibinfo{journal}{Phys. Rev. A} \textbf{\bibinfo{volume}{103}},
  \bibinfo{pages}{063112} (\bibinfo{year}{2021}).
	

\bibitem[{\citenamefont{Li et~al.}(2014)\citenamefont{Li, Geng, Liu, Deng,
  C.Wu, Peng, Gong, and Liu}}]{qorb1}
\bibinfo{author}{\bibfnamefont{M.}~\bibnamefont{Li}},
  \bibinfo{author}{\bibfnamefont{J.-W.} \bibnamefont{Geng}},
  \bibinfo{author}{\bibfnamefont{H.}~\bibnamefont{Liu}},
  \bibinfo{author}{\bibfnamefont{Y.}~\bibnamefont{Deng}},
  \bibinfo{author}{\bibnamefont{C.Wu}}, \bibinfo{author}{\bibfnamefont{L.-Y.}
  \bibnamefont{Peng}}, \bibinfo{author}{\bibfnamefont{Q.}~\bibnamefont{Gong}},
  \bibnamefont{and} \bibinfo{author}{\bibfnamefont{Y.}~\bibnamefont{Liu}},
  \emph{\bibinfo{title}{Classical-quantum correspondence for above-threshold
  ionization}}, \bibinfo{journal}{Phys.~Rev.~Lett.}
  \textbf{\bibinfo{volume}{112}}, \bibinfo{pages}{113002}
  (\bibinfo{year}{2014}).

\bibitem[{\citenamefont{Fang et~al.}(2019)\citenamefont{Fang, He, Han, Ge, Yu,
  Ma, Deng, and Liu}}]{qtmc2}
\bibinfo{author}{\bibfnamefont{Y.}~\bibnamefont{Fang}},
  \bibinfo{author}{\bibfnamefont{C.}~\bibnamefont{He}},
  \bibinfo{author}{\bibfnamefont{M.}~\bibnamefont{Han}},
  \bibinfo{author}{\bibfnamefont{P.}~\bibnamefont{Ge}},
  \bibinfo{author}{\bibfnamefont{X.}~\bibnamefont{Yu}},
  \bibinfo{author}{\bibfnamefont{X.}~\bibnamefont{Ma}},
  \bibinfo{author}{\bibfnamefont{Y.}~\bibnamefont{Deng}}, \bibnamefont{and}
  \bibinfo{author}{\bibfnamefont{Y.}~\bibnamefont{Liu}},
  \emph{\bibinfo{title}{Strong-field ionization of ar atoms with a
  ${45}^{\ensuremath{\circ}}$ cross-linearly-polarized two-color laser field}},
  \bibinfo{journal}{Phys. Rev. A} \textbf{\bibinfo{volume}{100}},
  \bibinfo{pages}{013414} (\bibinfo{year}{2019}).

\bibitem[{\citenamefont{Shvetsov-Shilovski
  et~al.}(2016)\citenamefont{Shvetsov-Shilovski, Lein, Madsen, R\"as\"anen,
  Lemell, Burgd\"orfer, Arb\'o, and T\"ok\'esi}}]{s2step}
\bibinfo{author}{\bibfnamefont{N.~I.} \bibnamefont{Shvetsov-Shilovski}},
  \bibinfo{author}{\bibfnamefont{M.}~\bibnamefont{Lein}},
  \bibinfo{author}{\bibfnamefont{L.~B.} \bibnamefont{Madsen}},
  \bibinfo{author}{\bibfnamefont{E.}~\bibnamefont{R\"as\"anen}},
  \bibinfo{author}{\bibfnamefont{C.}~\bibnamefont{Lemell}},
  \bibinfo{author}{\bibfnamefont{J.}~\bibnamefont{Burgd\"orfer}},
  \bibinfo{author}{\bibfnamefont{D.~G.} \bibnamefont{Arb\'o}},
  \bibnamefont{and}
  \bibinfo{author}{\bibfnamefont{K.}~\bibnamefont{T\"ok\'esi}},
  \emph{\bibinfo{title}{Semiclassical two-step model for strong-field
  ionization}}, \bibinfo{journal}{Phys. Rev. A} \textbf{\bibinfo{volume}{94}},
  \bibinfo{pages}{013415} (\bibinfo{year}{2016}).

\bibitem[{\citenamefont{Popruzhenko}(2014{\natexlab{a}})}]{tunr2}
\bibinfo{author}{\bibfnamefont{S.~V.} \bibnamefont{Popruzhenko}},
  \emph{\bibinfo{title}{Keldysh theory of strong field ionization: history,
  applications, difficulties and perspectives}}, \bibinfo{journal}{Journal of
  Physics B: Atomic, Molecular and Optical Physics}
  \textbf{\bibinfo{volume}{47}}(\bibinfo{number}{20}), \bibinfo{pages}{204001}
  (\bibinfo{year}{2014}{\natexlab{a}}).

\bibitem[{\citenamefont{Becker et~al.}(2002)\citenamefont{Becker, Grasbon,
  Kopold, Milo\u{s}evi\'c, Paulus, and Walther}}]{becker1}
\bibinfo{author}{\bibfnamefont{W.}~\bibnamefont{Becker}},
  \bibinfo{author}{\bibfnamefont{F.}~\bibnamefont{Grasbon}},
  \bibinfo{author}{\bibfnamefont{R.}~\bibnamefont{Kopold}},
  \bibinfo{author}{\bibfnamefont{D.~B.} \bibnamefont{Milo\u{s}evi\'c}},
  \bibinfo{author}{\bibfnamefont{G.~G.} \bibnamefont{Paulus}},
  \bibnamefont{and} \bibinfo{author}{\bibfnamefont{H.}~\bibnamefont{Walther}},
  \emph{\bibinfo{title}{Above-threshold ionization: From classical features to
  quantum effects}}, \bibinfo{journal}{Adv. At. Mol. Opt. Phys.}
  \textbf{\bibinfo{volume}{48}}, \bibinfo{pages}{35} (\bibinfo{year}{2002}).

\bibitem[{\citenamefont{Popov}(2005)}]{itm1}
\bibinfo{author}{\bibfnamefont{V.~S.} \bibnamefont{Popov}},
  \emph{\bibinfo{title}{Imaginary-time method in quantum mechanics and field
  theory}}, \bibinfo{journal}{Physics of Atomic Nuclei}
  \textbf{\bibinfo{volume}{68}}, \bibinfo{pages}{686} (\bibinfo{year}{2005}).

\bibitem[{\citenamefont{Popruzhenko}(2014{\natexlab{b}})}]{itm4}
\bibinfo{author}{\bibfnamefont{S.~V.} \bibnamefont{Popruzhenko}},
  \emph{\bibinfo{title}{Invariant form of coulomb corrections in the theory of
  nonlinear ionization of atoms by intense laser radiation}},
  \bibinfo{journal}{JETP} \textbf{\bibinfo{volume}{145}}, \bibinfo{pages}{580}
  (\bibinfo{year}{2014}{\natexlab{b}}).

\bibitem[{\citenamefont{Lai et~al.}(2015)\citenamefont{Lai, Poli, Schomerus,
  and Faria}}]{xv7}
\bibinfo{author}{\bibfnamefont{X.-Y.} \bibnamefont{Lai}},
  \bibinfo{author}{\bibfnamefont{C.}~\bibnamefont{Poli}},
  \bibinfo{author}{\bibfnamefont{H.}~\bibnamefont{Schomerus}},
  \bibnamefont{and} \bibinfo{author}{\bibfnamefont{C.~F. d.~M.}
  \bibnamefont{Faria}}, \emph{\bibinfo{title}{Influence of the coulomb
  potential on above-threshold ionization: A quantum-orbit analysis beyond the
  strong-field approximation}}, \bibinfo{journal}{Phys. Rev. A}
  \textbf{\bibinfo{volume}{92}}, \bibinfo{pages}{043407}
  (\bibinfo{year}{2015}).

\bibitem[{\citenamefont{Lai et~al.}(2017)\citenamefont{Lai, Yu, Huang, Hua,
  Gong, Quan, Faria, and Liu}}]{xv8}
\bibinfo{author}{\bibfnamefont{X.~Y.}~\bibnamefont{Lai}},
  \bibinfo{author}{\bibfnamefont{S.~G.}~\bibnamefont{Yu}},
  \bibinfo{author}{\bibfnamefont{Y.~Y.}~\bibnamefont{Huang}},
  \bibinfo{author}{\bibfnamefont{L.~Q.}~\bibnamefont{Hua}},
  \bibinfo{author}{\bibfnamefont{C.}~\bibnamefont{Gong}},
  \bibinfo{author}{\bibfnamefont{W.}~\bibnamefont{Quan}},
  \bibinfo{author}{\bibfnamefont{C.~F.} \bibnamefont{Faria}},
  \bibnamefont{and} \bibinfo{author}{\bibfnamefont{X.~J.}~\bibnamefont{Liu}},
  \emph{\bibinfo{title}{Near-threshold photoelectron holography beyond the
  strong-field approximation}}, \bibinfo{journal}{Phys. Rev. A}
  \textbf{\bibinfo{volume}{96}}, \bibinfo{pages}{013414}
  (\bibinfo{year}{2017}).


\bibitem[{\citenamefont{Sali\`{e}res et~al.}(2001)\citenamefont{Sali\`{e}res,
  Carr\'{e}, {Le D\'{e}roff}, Grasbon, Paulus, Walther, Kopold, Becker,
  Milo\u{s}evi\'c, Sanpera et~al.}}]{fein_ion}
\bibinfo{author}{\bibfnamefont{P.}~\bibnamefont{Sali\`{e}res}},
  \bibinfo{author}{\bibfnamefont{B.}~\bibnamefont{Carr\'{e}}},
  \bibinfo{author}{\bibfnamefont{L.}~\bibnamefont{{Le D\'{e}roff}}},
  \bibinfo{author}{\bibfnamefont{F.}~\bibnamefont{Grasbon}},
  \bibinfo{author}{\bibfnamefont{G.~G.} \bibnamefont{Paulus}},
  \bibinfo{author}{\bibfnamefont{H.}~\bibnamefont{Walther}},
  \bibinfo{author}{\bibfnamefont{R.}~\bibnamefont{Kopold}},
  \bibinfo{author}{\bibfnamefont{W.}~\bibnamefont{Becker}},
  \bibinfo{author}{\bibfnamefont{D.~B.} \bibnamefont{Milo\u{s}evi\'c}},
  \bibinfo{author}{\bibfnamefont{A.}~\bibnamefont{Sanpera}},
  \bibnamefont{et~al.}, \emph{\bibinfo{title}{Feynman's path-integral approach
  for intense-laser-atom interactions}}, \bibinfo{journal}{Science}
  \textbf{\bibinfo{volume}{292}}, \bibinfo{pages}{902} (\bibinfo{year}{2001}).

\bibitem[{\citenamefont{de~Morisson~Faria and Maxwell}(2020)}]{qtreview}
\bibinfo{author}{\bibfnamefont{C.~F.} \bibnamefont{de~Morisson~Faria}}
  \bibnamefont{and} \bibinfo{author}{\bibfnamefont{A.~S.}
  \bibnamefont{Maxwell}}, \emph{\bibinfo{title}{It is all about phases:
  ultrafast holographic photoelectron imaging}}, \bibinfo{journal}{Rep. Prog.
  Phys.} \textbf{\bibinfo{volume}{83}}, \bibinfo{pages}{034401}
  (\bibinfo{year}{2020}).

\bibitem[{\citenamefont{Lohr et~al.}(1997)\citenamefont{Lohr, Kleber, Kopold,
  and Becker}}]{sfa_sem1}
\bibinfo{author}{\bibfnamefont{A.}~\bibnamefont{Lohr}},
  \bibinfo{author}{\bibfnamefont{M.}~\bibnamefont{Kleber}},
  \bibinfo{author}{\bibfnamefont{R.}~\bibnamefont{Kopold}}, \bibnamefont{and}
  \bibinfo{author}{\bibfnamefont{W.}~\bibnamefont{Becker}},
  \emph{\bibinfo{title}{Above-threshold ionization in the tunneling regime}},
  \bibinfo{journal}{Phys. Rev. A} \textbf{\bibinfo{volume}{55}},
  \bibinfo{pages}{R4003} (\bibinfo{year}{1997}).

\bibitem[{\citenamefont{Kopold et~al.}(2000)\citenamefont{Kopold, Becker, and
  Kleber}}]{sfa_sem2}
\bibinfo{author}{\bibfnamefont{R.}~\bibnamefont{Kopold}},
  \bibinfo{author}{\bibfnamefont{W.}~\bibnamefont{Becker}}, \bibnamefont{and}
  \bibinfo{author}{\bibfnamefont{M.}~\bibnamefont{Kleber}},
  \emph{\bibinfo{title}{Quantum path analysis of high-order above-threshold
  ionization}}, \bibinfo{journal}{Optics Communications}
  \textbf{\bibinfo{volume}{179}}, \bibinfo{pages}{39} (\bibinfo{year}{2000}).

\bibitem[{\citenamefont{Korneev et~al.}(2012)\citenamefont{Korneev,
  Popruzhenko, Goreslavski, Yan, Bauer, Becker, K\"ubel, Kling, R\"odel,
  W\"unsche et~al.}}]{carp}
\bibinfo{author}{\bibfnamefont{P.~A.} \bibnamefont{Korneev}},
  \bibinfo{author}{\bibfnamefont{S.~V.} \bibnamefont{Popruzhenko}},
  \bibinfo{author}{\bibfnamefont{S.~P.} \bibnamefont{Goreslavski}},
  \bibinfo{author}{\bibfnamefont{T.-M.} \bibnamefont{Yan}},
  \bibinfo{author}{\bibfnamefont{D.}~\bibnamefont{Bauer}},
  \bibinfo{author}{\bibfnamefont{W.}~\bibnamefont{Becker}},
  \bibinfo{author}{\bibfnamefont{M.}~\bibnamefont{K\"ubel}},
  \bibinfo{author}{\bibfnamefont{M.~F.} \bibnamefont{Kling}},
  \bibinfo{author}{\bibfnamefont{C.}~\bibnamefont{R\"odel}},
  \bibinfo{author}{\bibfnamefont{M.}~\bibnamefont{W\"unsche}},
  \bibnamefont{et~al.}, \emph{\bibinfo{title}{Interference carpets in
  above-threshold ionization: From the coulomb-free to the coulomb-dominated
  regime}}, \bibinfo{journal}{Phys. Rev. Lett.} \textbf{\bibinfo{volume}{108}},
  \bibinfo{pages}{223601} (\bibinfo{year}{2012}).

\bibitem[{\citenamefont{Popruzhenko and Bauer}(2008)}]{coul1}
\bibinfo{author}{\bibfnamefont{S.}~\bibnamefont{Popruzhenko}} \bibnamefont{and}
  \bibinfo{author}{\bibfnamefont{D.}~\bibnamefont{Bauer}},
  \emph{\bibinfo{title}{Strong field approximation for systems with coulomb
  interaction}}, \bibinfo{journal}{Journal of Modern Optics}
  \textbf{\bibinfo{volume}{55}}(\bibinfo{number}{16}), \bibinfo{pages}{2573}
  (\bibinfo{year}{2008}).
	
	\bibitem[{\citenamefont{Yan et~al.}(2010)\citenamefont{Yan, Popruzhenko,
  Vrakking, and Bauer}}]{csfa3}
\bibinfo{author}{\bibfnamefont{T.-M.} \bibnamefont{Yan}},
  \bibinfo{author}{\bibfnamefont{S.~V.}~\bibnamefont{Popruzhenko}},
  \bibinfo{author}{\bibfnamefont{M.~J.~J.}~\bibnamefont{Vrakking}}, \bibnamefont{and}
  \bibinfo{author}{\bibfnamefont{D.}~\bibnamefont{Bauer}},
  \emph{\bibinfo{title}{Low-energy structures in strong field ionization
  revealed by quantum orbits}}, \bibinfo{journal}{Phys.~Rev.~Lett.}
  \textbf{\bibinfo{volume}{105}}, \bibinfo{pages}{253002}
  (\bibinfo{year}{2010}).


\bibitem[{\citenamefont{Li et~al.}(2016)\citenamefont{Li, Geng, Han, Liu, Peng,
  Gong, and Liu}}]{ccpati1}
\bibinfo{author}{\bibfnamefont{M.}~\bibnamefont{Li}},
  \bibinfo{author}{\bibfnamefont{J.-W.} \bibnamefont{Geng}},
  \bibinfo{author}{\bibfnamefont{M.}~\bibnamefont{Han}},
  \bibinfo{author}{\bibfnamefont{M.-M.} \bibnamefont{Liu}},
  \bibinfo{author}{\bibfnamefont{L.-Y.} \bibnamefont{Peng}},
  \bibinfo{author}{\bibfnamefont{Q.}~\bibnamefont{Gong}}, \bibnamefont{and}
  \bibinfo{author}{\bibfnamefont{Y.}~\bibnamefont{Liu}},
  \emph{\bibinfo{title}{Subcycle nonadiabatic strong-field tunneling
  ionization}}, \bibinfo{journal}{Phys. Rev. A} \textbf{\bibinfo{volume}{93}},
  \bibinfo{pages}{013402} (\bibinfo{year}{2016}).

\bibitem[{\citenamefont{Tulsky and Bauer}(2020)}]{tulsky}
\bibinfo{author}{\bibfnamefont{V.~A.} \bibnamefont{Tulsky}} \bibnamefont{and}
  \bibinfo{author}{\bibfnamefont{D.}~\bibnamefont{Bauer}},
  \emph{\bibinfo{title}{Numerical time-of-flight analysis of the strong-field
  photoeffect}}, \bibinfo{journal}{Phys. Rev. Res.}
  \textbf{\bibinfo{volume}{2}}, \bibinfo{pages}{043083} (\bibinfo{year}{2020}).

\bibitem[{\citenamefont{Tao and Scrinzi}(2012)}]{tserf}
\bibinfo{author}{\bibfnamefont{L.}~\bibnamefont{Tao}} \bibnamefont{and}
  \bibinfo{author}{\bibfnamefont{A.}~\bibnamefont{Scrinzi}},
  \emph{\bibinfo{title}{Photo-electron momentum spectra from minimal volumes:
  the time-dependent surface flux method}}, \bibinfo{journal}{New J. Phys.}
  \textbf{\bibinfo{volume}{14}}, \bibinfo{pages}{013021}
  (\bibinfo{year}{2012}).

\bibitem[{\citenamefont{Bohm}(1952)}]{bohm}
\bibinfo{author}{\bibfnamefont{D.}~\bibnamefont{Bohm}}, \emph{\bibinfo{title}{A
  suggested interpretation of the quantum theory in terms of "hidden"
  variables}}, \bibinfo{journal}{Physical Review}
  \textbf{\bibinfo{volume}{85}}, \bibinfo{pages}{166} (\bibinfo{year}{1952}).

\bibitem[{\citenamefont{Bell}(1987)}]{bell}
\bibinfo{author}{\bibfnamefont{J.~S.} \bibnamefont{Bell}},
  \emph{\bibinfo{title}{Speakable and Unspeakable in Quantum Mechanics}}
  (\bibinfo{publisher}{Cambridge University Press},
  \bibinfo{address}{Cambridge}, \bibinfo{year}{1987}).

\bibitem[{\citenamefont{Botheron and Pons}(2010)}]{bohm1DH}
\bibinfo{author}{\bibfnamefont{P.}~\bibnamefont{Botheron}} \bibnamefont{and}
  \bibinfo{author}{\bibfnamefont{B.}~\bibnamefont{Pons}},
  \emph{\bibinfo{title}{Self-consistent bohmian description of strong
  field-driven electron dynamics}}, \bibinfo{journal}{Phys.~Rev.~A}
  \textbf{\bibinfo{volume}{82}}, \bibinfo{pages}{021404(R)}
  (\bibinfo{year}{2010}).

\bibitem[{\citenamefont{Jooya et~al.}(2015{\natexlab{a}})\citenamefont{Jooya,
  Telnov, Li, and Chu}}]{bohm3Dsubsycle}
\bibinfo{author}{\bibfnamefont{H.~Z.} \bibnamefont{Jooya}},
  \bibinfo{author}{\bibfnamefont{D.~A.} \bibnamefont{Telnov}},
  \bibinfo{author}{\bibfnamefont{P.-C.} \bibnamefont{Li}}, \bibnamefont{and}
  \bibinfo{author}{\bibfnamefont{S.-I.} \bibnamefont{Chu}},
  \emph{\bibinfo{title}{Exploration of the subcycle multiphoton ionization
  dynamics and transient electron density structures with bohmian
  trajectories}}, \bibinfo{journal}{Phys.~Rev.~A}
  \textbf{\bibinfo{volume}{91}}, \bibinfo{pages}{063412}
  (\bibinfo{year}{2015}{\natexlab{a}}).

\bibitem[{\citenamefont{Sawada et~al.}(2014)\citenamefont{Sawada, Sato, and
  Ishikawa}}]{bohm1DH2}
\bibinfo{author}{\bibfnamefont{R.}~\bibnamefont{Sawada}},
  \bibinfo{author}{\bibfnamefont{T.}~\bibnamefont{Sato}}, \bibnamefont{and}
  \bibinfo{author}{\bibfnamefont{K.~L.} \bibnamefont{Ishikawa}},
  \emph{\bibinfo{title}{Analysis of strong-field enhanced ionization of
  molecules using bohmian trajectories}}, \bibinfo{journal}{Phys.~Rev.~A}
  \textbf{\bibinfo{volume}{90}}, \bibinfo{pages}{023404}
  (\bibinfo{year}{2014}).
	
	\bibitem[{\citenamefont{Wu et~al.}(2013{\natexlab{a}})\citenamefont{Wu,
  Augstein, and de~Morisson~Faria}}]{bohm1Dhhg}
\bibinfo{author}{\bibfnamefont{J.}~\bibnamefont{Wu}},
  \bibinfo{author}{\bibfnamefont{B.~B.} \bibnamefont{Augstein}},
  \bibnamefont{and} \bibinfo{author}{\bibfnamefont{C.}
  \bibnamefont{Figueira~de~Morisson~Faria}}, \emph{\bibinfo{title}{Bohmian-trajectory
  analysis of high-order-harmonic generation: Ensemble averages, nonlocality,
  and quantitative aspects}}, \bibinfo{journal}{Phys.~Rev.~A}
  \textbf{\bibinfo{volume}{88}}, \bibinfo{pages}{063416}
  (\bibinfo{year}{2013}{\natexlab{a}}).


\bibitem[{\citenamefont{Wu et~al.}(2013{\natexlab{b}})\citenamefont{Wu,
  Augstein, and Figueira~de Morisson~Faria}}]{bhhg2}
\bibinfo{author}{\bibfnamefont{J.}~\bibnamefont{Wu}},
  \bibinfo{author}{\bibfnamefont{B.~B.} \bibnamefont{Augstein}},
  \bibnamefont{and} \bibinfo{author}{\bibfnamefont{C.}~\bibnamefont{Figueira~de
  Morisson~Faria}}, \emph{\bibinfo{title}{Local dynamics in high-order-harmonic
  generation using bohmian trajectories}}, \bibinfo{journal}{Phys. Rev. A}
  \textbf{\bibinfo{volume}{88}}, \bibinfo{pages}{023415}
  (\bibinfo{year}{2013}{\natexlab{b}}).

\bibitem[{\citenamefont{Jooya et~al.}(2015{\natexlab{b}})\citenamefont{Jooya,
  Telnov, Li, and Chu}}]{bohm3Dhhg}
\bibinfo{author}{\bibfnamefont{H.~Z.} \bibnamefont{Jooya}},
  \bibinfo{author}{\bibfnamefont{D.~A.} \bibnamefont{Telnov}},
  \bibinfo{author}{\bibfnamefont{P.-C.} \bibnamefont{Li}}, \bibnamefont{and}
  \bibinfo{author}{\bibfnamefont{S.-I.} \bibnamefont{Chu}},
  \emph{\bibinfo{title}{Investigation of the characteristic properties of
  high-order harmonic spectrum in atoms using bohmian trajectories}},
  \bibinfo{journal}{J.~Phys.~B} \textbf{\bibinfo{volume}{48}},
  \bibinfo{pages}{195401} (\bibinfo{year}{2015}{\natexlab{b}}).

\bibitem[{\citenamefont{Zimmermann et~al.}(2016)\citenamefont{Zimmermann,
  Mishra, Doran, Gordon, and Landsman}}]{landsman_bohm}
\bibinfo{author}{\bibfnamefont{T.}~\bibnamefont{Zimmermann}},
  \bibinfo{author}{\bibfnamefont{S.}~\bibnamefont{Mishra}},
  \bibinfo{author}{\bibfnamefont{B.~R.} \bibnamefont{Doran}},
  \bibinfo{author}{\bibfnamefont{D.~F.} \bibnamefont{Gordon}},
  \bibnamefont{and} \bibinfo{author}{\bibfnamefont{A.~S.}
  \bibnamefont{Landsman}}, \emph{\bibinfo{title}{Tunneling time and weak
  measurement in strong field ionization}}, \bibinfo{journal}{Phys. Rev. Lett.}
  \textbf{\bibinfo{volume}{116}}, \bibinfo{pages}{233603}
  (\bibinfo{year}{2016}).

\bibitem[{\citenamefont{I.A.Ivanov et~al.}(2017)\citenamefont{I.A.Ivanov, Nam,
  and Kim}}]{bomii}
\bibinfo{author}{\bibnamefont{I.A.Ivanov}},
  \bibinfo{author}{\bibfnamefont{C.~H.} \bibnamefont{Nam}}, \bibnamefont{and}
  \bibinfo{author}{\bibfnamefont{K.~T.} \bibnamefont{Kim}},
  \emph{\bibinfo{title}{Exit point in the strong field ionization process}},
  \bibinfo{journal}{Scientific Reports} \textbf{\bibinfo{volume}{7}},
  \bibinfo{pages}{39919} (\bibinfo{year}{2017}).

\bibitem[{\citenamefont{Ivanov and Kim}(2022{\natexlab{a}})}]{cori1}
\bibinfo{author}{\bibfnamefont{I.}~\bibnamefont{Ivanov}} \bibnamefont{and}
  \bibinfo{author}{\bibfnamefont{K.~T.} \bibnamefont{Kim}},
  \emph{\bibinfo{title}{Analysis of correlations in strong field ionization}},
  \bibinfo{journal}{J.~Phys.~B} \textbf{\bibinfo{volume}{55}},
  \bibinfo{pages}{055001} (\bibinfo{year}{2022}{\natexlab{a}}).

\bibitem[{\citenamefont{Ivanov and Kim}(2022{\natexlab{b}})}]{cori2}
\bibinfo{author}{\bibfnamefont{I.}~\bibnamefont{Ivanov}} \bibnamefont{and}
  \bibinfo{author}{\bibfnamefont{K.~T.} \bibnamefont{Kim}},
  \emph{\bibinfo{title}{Joint probability calculation of the lateral velocity
  distribution in strong field ionization process.}}, \bibinfo{journal}{Sci.
  Rep.} \textbf{\bibinfo{volume}{12}}, \bibinfo{pages}{19533}
  (\bibinfo{year}{2022}{\natexlab{b}}).

\bibitem[{\citenamefont{Ivanov et~al.}(2023)\citenamefont{Ivanov, Kheifets, and
  Kim}}]{cori3}
\bibinfo{author}{\bibfnamefont{I.~A.} \bibnamefont{Ivanov}},
  \bibinfo{author}{\bibfnamefont{A.~S.} \bibnamefont{Kheifets}},
  \bibnamefont{and} \bibinfo{author}{\bibfnamefont{K.~T.} \bibnamefont{Kim}},
  \emph{\bibinfo{title}{Correlation analysis of frustrated tunneling
  ionization}}, \bibinfo{journal}{Phys.~Rev.~A} \textbf{\bibinfo{volume}{107}},
  \bibinfo{pages}{043106} (\bibinfo{year}{2023}).

\bibitem[{\citenamefont{Sarsa et~al.}(2003)\citenamefont{Sarsa, G\'{a}lvez, and
  Buendia}}]{oep1}
\bibinfo{author}{\bibfnamefont{A.}~\bibnamefont{Sarsa}},
  \bibinfo{author}{\bibfnamefont{F.~J.} \bibnamefont{G\'{a}lvez}},
  \bibnamefont{and} \bibinfo{author}{\bibfnamefont{E.}~\bibnamefont{Buendia}},
  \emph{\bibinfo{title}{A parametrized optimized effective potential for
  atoms}}, \bibinfo{journal}{J.~Phys.~B} \textbf{\bibinfo{volume}{36}},
  \bibinfo{pages}{4393} (\bibinfo{year}{2003}).

\bibitem[{\citenamefont{Lambropoulos and Petrosyan}(2007)}]{lampe}
\bibinfo{author}{\bibfnamefont{P.}~\bibnamefont{Lambropoulos}}
  \bibnamefont{and}
  \bibinfo{author}{\bibfnamefont{D.}~\bibnamefont{Petrosyan}},
  \emph{\bibinfo{title}{Fundamentals of Quantum Optics and Quantum
  Information}} (\bibinfo{publisher}{Springer-Verlag},
  \bibinfo{address}{Berlin}, \bibinfo{year}{2007}).

\bibitem[{\citenamefont{Ivanov}(2014)}]{cuspm}
\bibinfo{author}{\bibfnamefont{I.~A.} \bibnamefont{Ivanov}},
  \emph{\bibinfo{title}{Evolution of the transverse photoelectron-momentum
  distribution for atomic ionization driven by a laser pulse with varying
  ellipticity}}, \bibinfo{journal}{Phys. Rev. A} \textbf{\bibinfo{volume}{90}},
  \bibinfo{pages}{013418} (\bibinfo{year}{2014}).

\bibitem[{\citenamefont{Ivanov and Kheifets}(2013)}]{circ6}
\bibinfo{author}{\bibfnamefont{I.~A.} \bibnamefont{Ivanov}} \bibnamefont{and}
  \bibinfo{author}{\bibfnamefont{A.~S.} \bibnamefont{Kheifets}},
  \emph{\bibinfo{title}{Time delay in atomic photoionization with circularly
  polarized light}}, \bibinfo{journal}{Phys. Rev. A}
  \textbf{\bibinfo{volume}{87}}, \bibinfo{pages}{033407}
  (\bibinfo{year}{2013}).

\bibitem[{\citenamefont{I.A.Ivanov et~al.}(2016)\citenamefont{I.A.Ivanov,
  Dubau, and Kim}}]{ndim}
\bibinfo{author}{\bibnamefont{I.A.Ivanov}},
  \bibinfo{author}{\bibfnamefont{J.}~\bibnamefont{Dubau}}, \bibnamefont{and}
  \bibinfo{author}{\bibfnamefont{K.~T.} \bibnamefont{Kim}},
  \emph{\bibinfo{title}{Nondipole effects in strong-field ionization}},
  \bibinfo{journal}{Phys.~Rev.~A} \textbf{\bibinfo{volume}{94}},
  \bibinfo{pages}{033405} (\bibinfo{year}{2016}).

\bibitem[{\citenamefont{I.A.Ivanov and Y.K.Ho}(1999)}]{ykh1}
\bibinfo{author}{\bibnamefont{I.A.Ivanov}} \bibnamefont{and}
  \bibinfo{author}{\bibnamefont{Y.K.Ho}},
  \emph{\bibinfo{title}{High-angular-momentum ($ {L} > 3$ ) doubly excited
  resonance states of positronium negative ion}},
  \bibinfo{journal}{Phys.~Rev.~A} \textbf{\bibinfo{volume}{60}},
  \bibinfo{pages}{1015} (\bibinfo{year}{1999}).

\bibitem[{\citenamefont{Bader et~al.}(2013)\citenamefont{Bader, Blanes, and
  Casas}}]{itp}
\bibinfo{author}{\bibfnamefont{P.}~\bibnamefont{Bader}},
  \bibinfo{author}{\bibfnamefont{S.}~\bibnamefont{Blanes}}, \bibnamefont{and}
  \bibinfo{author}{\bibfnamefont{F.}~\bibnamefont{Casas}},
  \emph{\bibinfo{title}{Solving the schrodinger eigenvalue problem by the
  imaginary time propagation technique using splitting methods with complex
  coefficients}}, \bibinfo{journal}{Journal of Chemical Physics}
  \textbf{\bibinfo{volume}{139}}, \bibinfo{pages}{124117}
  (\bibinfo{year}{2013}).

\bibitem[{\citenamefont{Nurhuda and Faisal}(1999)}]{velocity1}
\bibinfo{author}{\bibfnamefont{M.}~\bibnamefont{Nurhuda}} \bibnamefont{and}
  \bibinfo{author}{\bibfnamefont{F.~H.~M.} \bibnamefont{Faisal}},
  \emph{\bibinfo{title}{Numerical solution of time-dependent schr\"odinger
  equation for multiphoton processes: A matrix iterative method}},
  \bibinfo{journal}{Phys. Rev. A}
  \textbf{\bibinfo{volume}{60}}(\bibinfo{number}{4}), \bibinfo{pages}{3125}
  (\bibinfo{year}{1999}).

\bibitem[{\citenamefont{Sun et~al.}(2014)\citenamefont{Sun, Li, Yu, Deng, Gong,
  and Liu1}}]{sun1}
\bibinfo{author}{\bibfnamefont{X.}~\bibnamefont{Sun}},
  \bibinfo{author}{\bibfnamefont{M.}~\bibnamefont{Li}},
  \bibinfo{author}{\bibfnamefont{J.}~\bibnamefont{Yu}},
  \bibinfo{author}{\bibfnamefont{Y.}~\bibnamefont{Deng}},
  \bibinfo{author}{\bibfnamefont{Q.}~\bibnamefont{Gong}}, \bibnamefont{and}
  \bibinfo{author}{\bibfnamefont{Y.}~\bibnamefont{Liu}},
  \emph{\bibinfo{title}{Calibration of the initial longitudinal momentum spread
  of tunneling ionization}}, \bibinfo{journal}{Phys.~Rev.~A}
  \textbf{\bibinfo{volume}{89}}, \bibinfo{pages}{045402}
  (\bibinfo{year}{2014}).


\end{thebibliography}

\end{document}